\documentclass[reprint,prx,superscriptaddress, amsmath,amssymb, aps,]{revtex4-2}
\usepackage[dvipsnames]{xcolor}
\usepackage{tcolorbox}
\usepackage{wrapfig}
\usepackage{floatrow}
\usepackage{graphicx}%
\usepackage[english]{babel}
\usepackage{graphicx}
\usepackage{nccmath}
\usepackage{float}
\usepackage[utf8]{inputenc}
\usepackage[colorlinks,allcolors=cyan!70!black]{hyperref}
\usepackage{blindtext}
\usepackage{physics,amsmath,nccmath}
\usepackage{mathtools}
\usepackage{lipsum}
\newlength{\arrow}
\newcommand{\rcomm}[1]{\textcolor{black}{#1}}
\newcommand*{\myrightarrow}[1]{\xrightarrow{\mathmakebox[\arrow]{#1}}}
\newcommand*{\myxrightleftharpoons}[2]{\xrightleftharpoons[{#1}]{\mathmakebox[\arrow]{#2}}}
\newlength{\extralongarrow}
\newcommand{\Cov}{\mathrm{Cov}}

\newcommand{\lb}{\langle}
\newcommand{\rb}{\rangle}

\begin{document}
\title{Characterizing the non-monotonic behavior of mutual information\\ along biochemical reaction cascades}
\author{Raymond Fan}
\affiliation{Department of Physics, University of Toronto, 60 St.~George St., Ontario M5S 1A7, Canada}
\affiliation{Department of Chemical \& Physical Sciences, University of Toronto, Mississauga, Ontario L5L 1C6, Canada}
\author{Andreas Hilfinger}
\email[andreas.hilfinger@utoronto.ca]{}
\affiliation{Department of Physics, University of Toronto, 60 St.~George St., Ontario M5S 1A7, Canada}
\affiliation{Department of Chemical \& Physical Sciences, University of Toronto, Mississauga, Ontario L5L 1C6, Canada}
\affiliation{Department of Cell \& Systems Biology, University of Toronto, 25 Harbord St, Toronto, Ontario M5S 3G5}
\affiliation{Department of Mathematics, University of Toronto, 40 St.~George St., Toronto, Ontario M5S 2E4}

\begin{abstract}
Cells sense environmental signals and transmit information intracellularly through changes in the abundance of molecular components. Such molecular abundances can be measured in single cells and exhibit significant heterogeneity in clonal populations even in identical environments. Experimentally observed joint probability distributions \rcomm{can} then \rcomm{be used to} quantify the covariability and mutual information between molecular abundances \rcomm{along signaling cascades}.
However, because stationary state abundances along stochastic biochemical reaction cascades are not conditionally independent, their mutual information is not constrained by the data processing inequality.  Here, we report the conditions under which the mutual information between stationary state abundances increases along a cascade of biochemical reactions. This non-monotonic behavior can be intuitively understood in terms of noise propagation and time-averaging stochastic fluctuations that are short-lived compared to an extrinsic signal.
Our results \rcomm{re-emphasize} that mutual information measurements of stationary state distributions of cellular components \rcomm{may be of limited utility for characterizing cellular signaling processes} because they do not measure information transfer.
\vspace{.35cm}
\end{abstract}

\maketitle

\section*{Introduction}
\vspace{-1.em}
Cells respond and adapt to changing environments by transmitting information through biochemical reaction networks. A quantitative framework for analyzing this process is information theory, which was originally developed in the context of telecommunications \cite{shannon} but has recently been applied to various biological problems such as determining the amount of information encoded in genes during fruit fly development \cite{dubuis2013positional} and deriving fundamental limits on the suppression of molecular fluctuations \cite{Lestas2010}.

In biochemical reaction networks, information is transmitted through the time-varying concentrations of molecules \cite{Moor2023, tostevin2009mutual, tostevin2010mutual, Ronde2010, mattingly2021escherichia}.
Mutual information between temporal trajectories obeys the data processing inequality which establishes how information can only be lost but never recovered along communication channels \cite{Cover2006}. This key theorem has been applied to derive constraints on noise suppression in cells \cite{Lestas2010} and a theoretical analysis of intracellular trajectories has characterized the  information transfer through biochemical network motifs \cite{tostevin2010mutual}.
However, for cellular pathways it is \rcomm{practically} impossible to directly estimate mutual information between trajectories from experimental data \cite{meijers2021behavior}.
Recent work has thus focused on estimating the mutual information between trajectories \rcomm{indirectly through model simulations} \cite{duso2019path, reinhardt2023path, Moor2023}.

In contrast, the probability distributions of molecular abundances across a population of cells are directly experimentally accessible and are commonly reported to summarize the non-genetic variability of molecular abundances \cite{Taniguchi2010, vogel2012insights, vistain2021single}. While such distributions are a powerful tool to analyze and describe stochastic fluctuations in cells \cite{Munsky2012}, the mutual information between variables in such distributions does not measure information transfer and is not constrained by the data processing inequality. The premise of the data processing inequality requires that a component becomes independent of an upstream signal when conditioned on an intermediate which is not the case for stationary state molecular levels in biochemical reaction cascades. Components in  reaction cascades only become independent when conditioned on the entire histories of the intermediates \cite{Hilfinger2011,bowsher2012identifying, bowsher2013fidelity}.

Although the premise of the theorem is not satisfied, the data processing inequality has been incorrectly stated or implied when discussing stationary state distributions \cite{Ziv2007, Biswas2016, Nandi2019, Roy2021}. To counter these claims, recent work has reported that the mutual information between a signal and downstream components increases in the special case of a biochemical cascade with components of equal lifetimes and increasing average abundances \cite{Pilkiewicz2016, Rowland2021}.

Here, we establish general conditions under which the mutual information between stationary distributions of components increases along simple biochemical cascades.
Combining exact numerical simulations over a wide range of parameters with analytical approximations, we show that in a three variable linear cascade both inequalities implied by the data processing inequality will be violated when a slow read-out component averages out fast fluctuations of a noisy intermediate that responds to a slow extrinsic signal.
This result contradicts the naive expectation that as long as the timescale of a signal is much slower than the timescales of the reaction network the mutual information between stationary state abundances decrease along biochemical cascades.
We further show that this non-monotonic behavior is not special to linear cascades but also occurs in more complex reaction systems such as kinetic proofreading.

\vspace{-.5cm}

\begin{figure*}[btp!]
\centering
\includegraphics[width=1\textwidth]{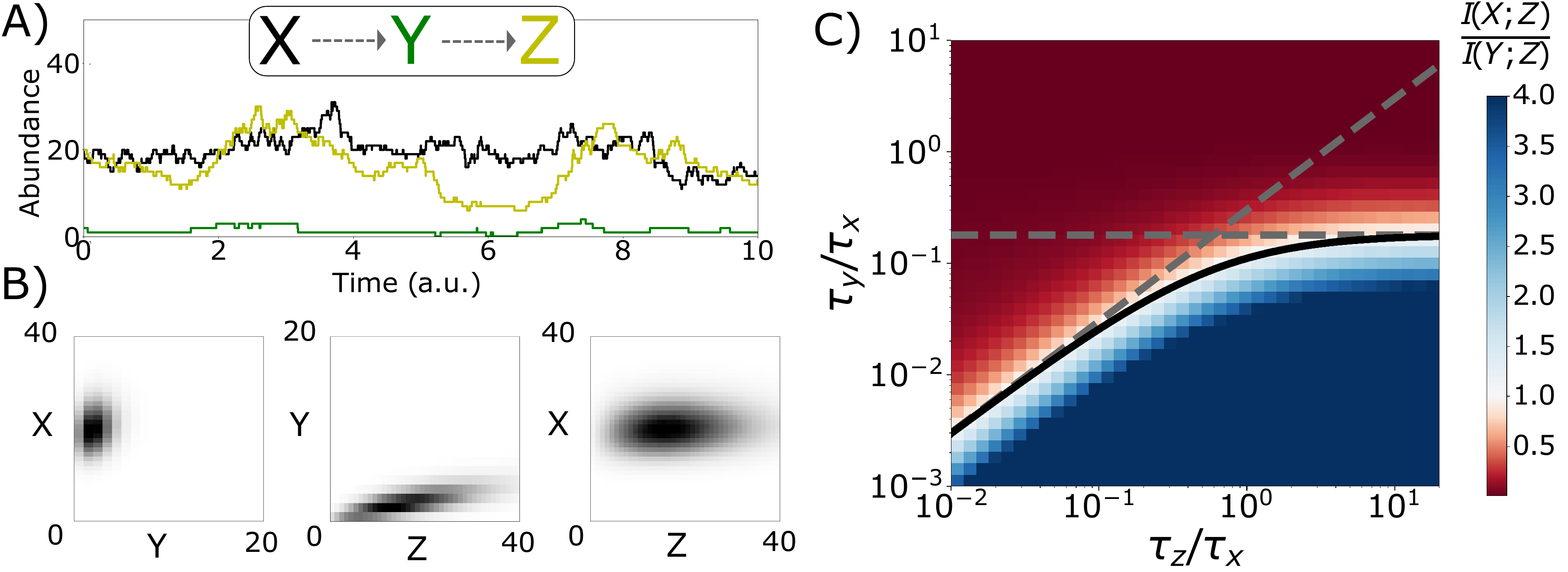}
\caption{\textbf{
The stationary state distributions of molecular abundances in biochemical cascades are not constrained by Eq.~\ref{EQ: DPI YZ XZ}.}
A) Example time trace of the linear cascade defined in Eq.~\ref{EQ: Two Step Cascade} in which a signal $X$ affects the production of $Y$, which in turn affects the production of $Z$. B) Corresponding stationary state distribution of molecular abundances. C) Exact numerical simulation results for the mutual information between pairs of variables in the stationary state distribution of the cascade defined by Eq.~\ref{EQ: Two Step Cascade}.
The blue region indicates the regime in which the stationary state abundance of the final read-out component exhibits a lower mutual information with the intermediate of the cascade than with the further removed upstream signal. The solid black line indicates the analytical condition of Eq.~\ref{EQ: XZ YZ Correlation Inequality} that approximately determines the boundary at which $I(X; Z) = I(Y; Z)$. Dashed grey lines indicate the approximate necessary (but not sufficient) conditions of Eqs.~\ref{EQ: XZ YZ Diagonal} and \ref{EQ: XZ YZ Inequality Maximum TyTx}.
The depicted results correspond to systems with average abundances $\langle x \rangle = \langle z \rangle = 20$ and $\langle y \rangle = 2$, representing a process with a noisy intermediate. Numerical simulations for other noise regimes show similar behavior, see supplement \cite{SI}.
}
\label{FIG: Intro Cartoon and first result}
\end{figure*}

\section*{Results}
\vspace{-.35cm}
\subsection*{Mutual information is non-monotonic along biochemical cascades}
\vspace{-.25cm}
To analyze mutual information in biochemical signaling we consider the following stochastic process in which three components $X,Y,Z$ form a linear cascade
\begin{equation}
\label{EQ: Two Step Cascade}
\begin{array}{rcl}
x\hspace{-3pt}&\myrightarrow{\lambda}&\hspace{-3pt} x + 1 \\
x\hspace{-3pt}&\myrightarrow{x /\tau_x}&\hspace{-3pt} x - 1 
\end{array}\hspace{8pt}
\begin{array}{rcl}
y\hspace{-3pt} & \myrightarrow{\alpha\hspace{.1em}x}& \hspace{-3pt}y + 1 \\
y\hspace{-3pt} & \myrightarrow{ y /\tau_y}& \hspace{-3pt}y - 1
\end{array}\hspace{8pt}
\begin{array}{rcl}
z\hspace{-3pt}& \myrightarrow{\beta y} & \hspace{-3pt}z + 1 \\
z\hspace{-3pt} & \myrightarrow{z /\tau_z} & \hspace{-3pt}z - 1 
\end{array}\hspace{3pt}.
\end{equation}
Here, variables linearly affect the production of the next downstream and each component undergoes first order degradation with respective average lifetimes $\tau_x, \tau_y, \tau_z$.

The covariances between components of this generic regulatory cascade have been previously reported \cite{Paulsson2005}. Here, we consider the joint stationary state distribution of molecular abundances $P_\text{ss}(x,y,z)$ and determine the mutual information \cite{shannon} between pairs of abundances, %
\begin{equation*}
I(X;Y) : = \sum_{x, y} P_\text{ss}(x, y)\log_2\left(\frac{P_\text{ss}(x, y)}{P_\text{ss}(x)P_\text{ss}(y)} \right) \hspace{3pt},
\end{equation*}
and analogously for any other pair.

The mutual information between pairs of components that form a Markov chain $X\to Y\to Z$ is constrained by the data processing inequality \cite{Cover2006} which implies that
\begin{align}
I(X; Z) & \leq I(Y; Z)
\label{EQ: DPI YZ XZ}\\
I(X; Z) & \leq I(X; Y) \hspace{3pt}.
\label{EQ: DPI XY XZ}
\end{align}
This pair of inequalities establishes that information can only be lost along Markov chains and once information is lost it cannot be recovered through processing. However, 
while the biochemical cascade defined in Eq.~\ref{EQ: Two Step Cascade} is Markovian in time, i.e., its future evolution depends only on the current state but not its history, the stationary state probability distributions of the process do not satisfy 
\mbox{$P(X,Y,Z) = P(X) P(Y|X) P(Z|Y)$} \cite{Hilfinger2011,bowsher2012identifying, bowsher2013fidelity} and are thus not constrained by the data processing inequality. 

While this lack of conditional independence might look surprising, and has in fact been mistakenly assumed in related processes \cite{Ziv2007, Biswas2016, Nandi2019, Roy2021}, it intuitively follows because Eq.~\ref{EQ: Two Step Cascade} defines is a dynamically varying stochastic process rather than a cascade of static random variables. In other words, the conditional probability distribution $P(Z|Y)$ is not entirely determined by how the dynamics of $Z$ depends on $Y$ but, e.g., depends on how long the system spends in each $Y$-abundance state which in turn depends on the abundance of $X$.
The trajectories of abundances %
thus become conditionally independent only when conditioned on the temporal \emph{history} of upstream variables \cite{borst1999information, Tostevin2010, de2012feed, Mancini2013}.

We first analyze under which conditions 
the mutual information of the stationary state distribution of molecular abundances along the above biochemical cascade 
does not obey Eq.~\ref{EQ: DPI YZ XZ}. Of course a proven theorem cannot be violated, but since the premise of the data processing inequality is not satisfied by the quantities under consideration they do not need to satisfy the theorem's conclusion.
Using exact numerical simulations of the system defined by Eq.~\ref{EQ: Two Step Cascade} 
using the Gillespie algorithm \cite{Gillespie1977} we find that for any tested molecular abundance levels, Eq.~\ref{EQ: DPI YZ XZ} was reversed in the regime when $\tau_z/\tau_x$ is large and $\tau_y/\tau_x$ is small as indicated by the blue region in Fig.~\ref{FIG: Intro Cartoon and first result}B. In this regime the abundance of the intermediate component $Y$ contains significantly less information about the abundance of the final read-out variable $Z$ than the more distantly connected upstream signal $X$.
See supplement \cite{SI} for numerical simulation results for abundances other than those depicted in Fig.~\ref{FIG: Intro Cartoon and first result}B.

The above timescale dependence can be intuitively understood through approximate analytical arguments. For multivariate Gaussian distributions, the mutual information between components is related to their correlation through \mbox{$I(X;Y)=-\log_2(1-\rho_{xy} ^{2})/2$}. When stationary state distributions are approximately Gaussian, we can thus estimate the mutual information between components from their correlations, which can be exactly determined from the system's chemical master equation, see Appendix \ref{APP: Two Step Cascade Correlations and Mutual Information}.

Following this approach yields the following (approximate) necessary and sufficient condition to reverse Eq.~\ref{EQ: DPI YZ XZ}. 
\begin{equation}
\begin{split}
\label{EQ: XZ YZ Correlation Inequality}
&1 + \frac{\tau_y}{\tau_z}\left(1 + \frac{\eta_{y}^\text{int}}{\eta_\mathrm{sig} }\left(1 + \frac{\tau_y}{\tau_x}\right)\right) \left(1 + \frac{\tau_z}{\tau_x} \right) 
\\ 
&\leq \left( 1 + \frac{\tau_y}{\tau_z} \right) 
\Biggl(\frac{\eta_{y}^\text{int}}{\eta_\mathrm{sig} } + \frac{1}{1 + \mfrac{\tau_y}{\tau_x}} \Biggr)^{1/2}
\end{split}\hspace{3pt},
\end{equation}
where we have introduced the normalized signal variability and intrinsic noise terms
\begin{equation}
\label{EQ: Intrinsic Noise Notation}
\eta_\mathrm{sig} := \frac{\mathrm{Var}(x)}{\langle x \rangle^2}, \quad \eta_{y}^\text{int} := \frac{1}{\langle y \rangle}, \quad \eta_{z}^\text{int} := \frac{1}{\langle z \rangle}.
\end{equation}
Here, and throughout angular brackets denote stationary state averages.

In Fig.~\ref{FIG: Intro Cartoon and first result}C, the black line indicates the boundary defined by Eq.~\ref{EQ: XZ YZ Correlation Inequality} which agrees well with the numerical solutions for the cascade defined in Eq.~\ref{EQ: Two Step Cascade} with abundances
$\langle x \rangle = \langle z \rangle = 20, \langle y \rangle = 2$ such that $\eta_{y}^\text{int} / \eta_\mathrm{sig} = 10$. Eq.~\ref{EQ: XZ YZ Correlation Inequality} also accurately describes systems with different variability ratios as long as the average abundances of all molecules are larger than one molecule \cite{SI}.

The analytical condition of Eq.~\ref{EQ: XZ YZ Correlation Inequality} can be intuitively understood in the regimes in which $\tau_z$ is much slower or much faster than $\tau_x$.
When $\tau_z \ll \tau_x$ the inequality asymptotically becomes 
\begin{equation}
\label{EQ: XZ YZ Diagonal}
\sqrt{1 + \frac{\eta_{y}^\text{int}}{\eta_\mathrm{sig} }} 
\leq \frac{\tau_z}{\tau_y} ,
\end{equation} 
as long as $\tau_y \ll \tau_x$, see Appendix \ref{APP: Mutual Information Inequalities}. Eq.~\ref{EQ: XZ YZ Diagonal} is indicated by the diagonal line of Fig.~\ref{FIG: Intro Cartoon and first result}.

\begin{figure}[!hbt]
\vspace{-0.5em}
\centering
\includegraphics[width=0.85\columnwidth]{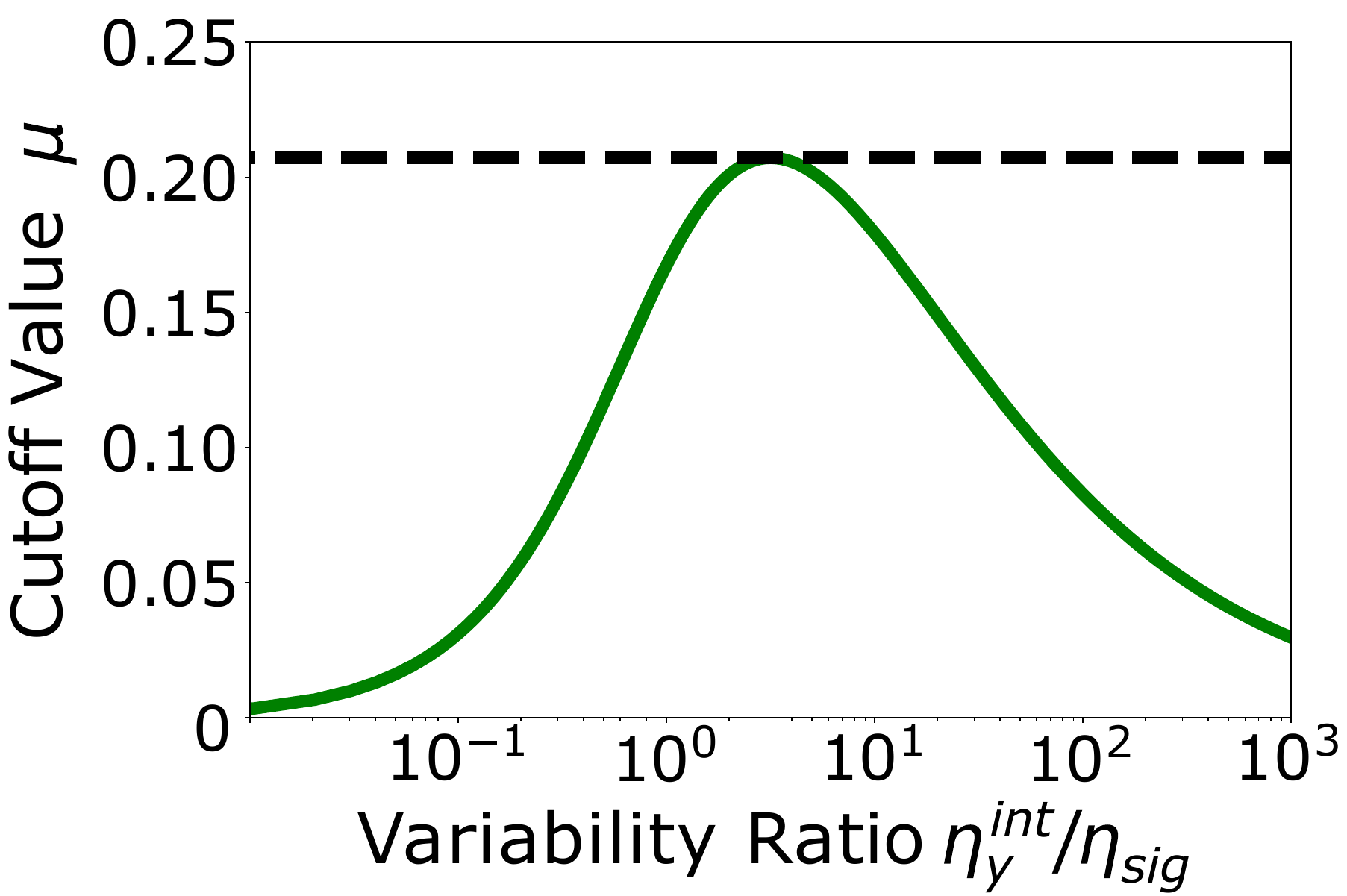}
\vspace{-0.5em}
\caption{\textbf{Analytical approximations predict that Eq.~\ref{EQ: DPI YZ XZ} will not be reversed in cascades with a slow intermediate.}
Eq.~\ref{EQ: XZ YZ Correlation Inequality} predicts that violating Eq.~\ref{EQ: DPI YZ XZ} requires
$\tau_y / \tau_x < \mu$ where the cutoff parameter $\mu$ depends on the variability ratio $\eta_{y}^\text{int}/\eta_\mathrm{sig} $ and approaches zero in both limits.
Dashed line indicates the analytically determined maximum 
of Eq.~\ref{EQ: XZ YZ Inequality Maximum TyTx}, which corresponds to the largest value of $\tau_y/\tau_x$ for which violations of Eq.~\ref{EQ: DPI YZ XZ} are expected. Regardless of any other system details we thus do not expect violations in cascades with a relatively slow intermediate component.}
\label{FIG:XZYZ Analysis}
\end{figure}

When $\tau_z \gg \tau_x$ the boundary asymptotically becomes a simple cutoff $\tau_y / \tau_x \lesssim \mu$ where $\mu$ is the root of a quintic polynomial in $\eta_{y}^\text{int}/\eta_\mathrm{sig} $, see Appendix \ref{APP: Mutual Information Inequalities}. While the root is not analytically accessible we can calculate it numerically (see Fig.~\ref{FIG:XZYZ Analysis})
and analytically determine its maximum value given by
\begin{equation}
\label{EQ: XZ YZ Inequality Maximum TyTx}
\mu^* = \frac{\sqrt{2}-1}{2}\approx 0.2071 \hspace{3pt} ,
\end{equation}
which is attained when $\eta_\mathrm{sig}  = (6 - 2 \sqrt{2})\eta_{y}^\text{int}$, %
suggesting that violations of Eq.~\ref{EQ: DPI YZ XZ} will not be observed in systems in which the lifetime of the intermediate variable is larger than one fifth of the input signal regardless of noise levels and other timescales.

These asymptotic behaviors of Eq.~\ref{EQ: XZ YZ Correlation Inequality} translate into a simple pair of necessary conditions on the timescales as indicated by the grey dashed lines in Fig.~\ref{FIG: Intro Cartoon and first result}C. The mutual information between stationary state distributions of molecular abundances along a linear cascade is thus expected to violate Eq.~\ref{EQ: DPI YZ XZ} when a slowly varying signal affects a downstream component through a short-lived and noisy intermediate.

\begin{figure*}[!hbt]
\centering
\includegraphics[width=1\textwidth]{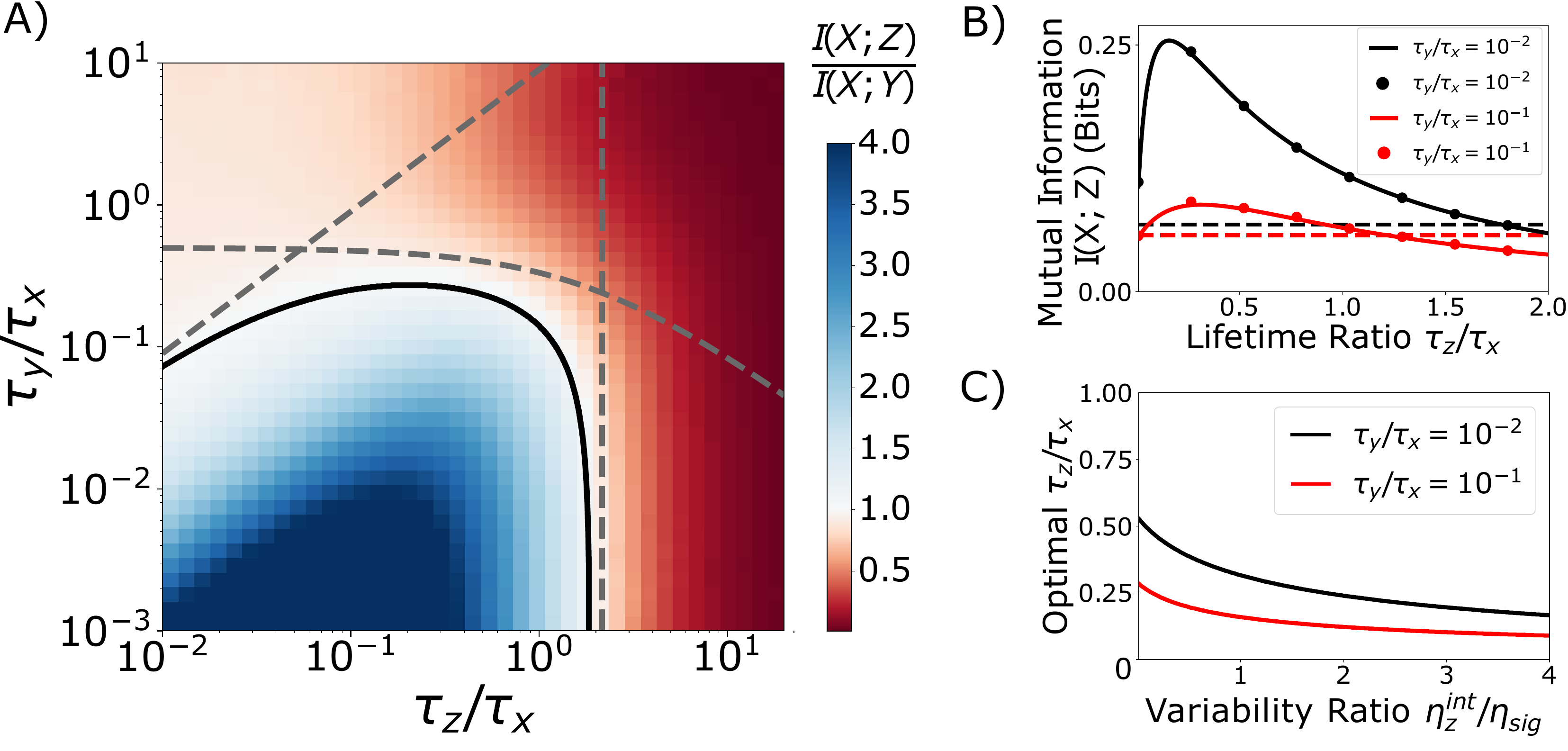}
\caption{\textbf{Mutual information between stationary state abundances can increase along a biochemical cascade.}
A) Exact numerical simulation results for the stationary state distributions of molecular abundances in the cascade defined by Eq.~\ref{EQ: Two Step Cascade} with noise parameters as in Fig.~\ref{FIG: Intro Cartoon and first result}.
When the signal timescale is slowest in the cascade, and $Y$ is fast compared to $Z$, the upstream signal exhibits a larger mutual information with the final read-out variable than with the intermediate as indicated by the blue region. The solid black line indicates the analytical approximation of Eq.~\ref{EQ: DPI XY XZ - Simplified Correlations} for the boundary at which $I(X; Z) = I(X; Y)$. Dashed grey lines indicate intuitively interpretable necessary (but not sufficient) conditions of Eqs.~\ref{EQ: Noise Limit Condition}--\ref{EQ: Simple Lifetime Condition} for violations of Eq.~\ref{EQ: DPI XY XZ} to occur. B) When keeping all other parameters fixed $I(X; Z)$ exhibits a maximum as a function of the relative timescale $\tau_z/\tau_x$. This maximum is significantly higher than $I(X; Y)$ (dashed lines) as long as the intermediate variable is sufficiently short-lived.
C) In the regime in which an optimal lifetime ratio exists, it decreases monotonically as the variability ratio $\eta_{z}^\text{int}/\eta_\mathrm{sig}$ increases, see Appendix \ref{APP: Mutual Information Two Step Appendix}.
}
\label{FIG:XYXZ Timescales}
\end{figure*}

The above results establish the conditions under which the mutual information between the last component and their upstream components can be non-monotonic, i.e., components that are further removed from the final read-out can have larger mutual information when compared to those closer connected to the final component. Next, we consider the complementary question when a seeming ``loss of information'' in the first step of the cascade will be recovered through an additional step, i.e., we analyze the conditions under which inequality Eq.~\ref{EQ: DPI XY XZ} is violated. 

Previous work \cite{Pilkiewicz2016, Rowland2021}  established that this inequality will be violated when the signaling timescale is long-lived compared to downstream components with equal lifetimes and increasing abundances along the cascade. Next, we generalize those results to biochemical cascades with arbitrary timescales and abundances.
For a given set of variability ratios the numerically observed behavior of the mutual information between stationary state abundances depends on the relative timescales of the system as indicated in Fig.~\ref{FIG:XYXZ Timescales}A. Changing the variability ratios by adjusting average abundances affects the possible behavior significantly. For example, when the variability of the read-out variable was comparable to the intrinsic noise of the intermediate, we did not observe violations of Eq.~\ref{EQ: DPI XY XZ} over the numerically explored timescales \cite{SI}.

To understand this dependence we derive approximate analytical conditions under which the mutual information between molecular abundances increases along biochemical cascades.
Following the same Gaussian approximation as above, we translate the exact (co)variance solutions for Eq.~\ref{EQ: Two Step Cascade} into the following approximate condition for processes to violate Eq.~\ref{EQ: DPI XY XZ}
\begin{align}
\frac{\left(1 + \mfrac{\tau_z}{\tau_x}\right)^2}{1 + \mfrac{\tau_y}{\tau_x}}&\left(\frac{1}{1 + \mfrac{\tau_z}{\tau_y}} + \frac{1}{1 + \mfrac{\tau_y}{\tau_z}} \frac{1}{1 + \mfrac{\tau_z}{\tau_x}}\right)
 +  \frac{\eta_{z}^\text{int}}{\eta_\mathrm{sig} }\left(1 + \frac{\tau_z}{\tau_x}\right)^2 \nonumber
 \\ 
 & \leq \frac{\eta_{y}^\text{int}}{\eta_\mathrm{sig} } \left(1 - \frac{\left(1 + \mfrac{\tau_z}{\tau_x}\right)^2}{1 + \mfrac{\tau_z}{\tau_y}}\right) + \frac{1}{1 + \mfrac{\tau_y}{\tau_x}} \hspace{3pt}.
 \label{EQ: DPI XY XZ - Simplified Correlations}
\end{align}

The boundary defined by Eq.~\ref{EQ: DPI XY XZ - Simplified Correlations} agrees well with the exact numerical results for a system with variability ratios $\eta_{y}^\text{int} / \eta_\mathrm{sig}  = 10$ and $\eta_{z}^\text{int} = \eta_\mathrm{sig} $, as indicated by the solid black line in Fig.~\ref{FIG:XYXZ Timescales}A. 

To intuitively understand the condition of Eq.~\ref{EQ: DPI XY XZ} we first consider the maximum $\tau_z / \tau_x$ above which violations become impossible. The exact cut-off value depends on both variability ratios (see supplement \cite{SI}) but is itself bounded by
\begin{equation}
\label{EQ: Noise Limit Condition}
\frac{\tau_z }{ \tau_x} \leq \sqrt{\frac{\eta_y^\text{int} }{ \eta_z^\text{int}}} - 1  \hspace{3pt},
\end{equation}
indicated by the vertical (dashed) grey line in Fig.~\ref{FIG:XYXZ Timescales}. Note, Eq.~\ref{EQ: Noise Limit Condition} explains why violations of Eq.~\ref{EQ: DPI XY XZ} were not observed in our numerical simulations when the read-out-variable was noisier than the intermediate 
\cite{SI}.

Considering the regime in which $\tau_z, \tau_y \ll \tau_x$, leads to the following necessary condition 
\begin{equation}
\label{EQ: Mutual Information Condition}
\frac{\tau_y}{\tau_z} \leq \mfrac{\eta_{y}^\text{int}}{\eta_{z}^\text{int}}-1 \hspace{3pt},
\end{equation}
which illustrates that the read-out variable $Z$ must become longer lived the closer its intrinsic noise gets to that of the intermediate in order for the mutual information to increase along the cascade.
The condition of Eq.~\ref{EQ: Mutual Information Condition} is indicated by the diagonal dashed grey line in Fig.~\ref{FIG:XYXZ Timescales}. 

Finally, we consider the limit in which the intermediate component is far noisier than the others, i.e., $\eta_y^\text{int} \gg \eta_\mathrm{sig} ,\eta_z^\text{int}$. In this regime, Eq.~\ref{EQ: DPI XY XZ - Simplified Correlations} becomes independent of the variability ratios and constrains the timescales through
\begin{equation}
\label{EQ: Simple Lifetime Condition}
\frac{\tau_y}{\tau_x} \leq \frac{1}{2}
\left(1 + \frac{\tau_z}{2\tau_x}\right)^{-1} \hspace{3pt}.
\end{equation}
indicated by the curved dashed grey line in Fig.~\ref{FIG:XYXZ Timescales}. Because the right hand side of Eq.~\ref{EQ: Simple Lifetime Condition} is bounded by $1/2$ it puts an upper limit on the lifetime of the intermediate variable for violations of Eq.~\ref{EQ: DPI XY XZ} to occur. This suggests that an increase in mutual information between molecular abundances will not be observed in biochemical cascades in which the lifetime of intermediate variable is longer than one half of that of the input signal.
Eqs.~\ref{EQ: Noise Limit Condition}, \ref{EQ: Mutual Information Condition}, and \ref{EQ: Simple Lifetime Condition} are derived in Appendix \ref{APP: Mutual Information Inequalities}.

While the above conditions are necessary but not sufficient they intuitively describe the qualitative behavior of the system: %
We conclude that the inequalities of Eqs.~\ref{EQ: DPI YZ XZ} and \ref{EQ: DPI XY XZ} are simultaneously reversed when the input signal $X$ is the slowest component, the intermediate component $Y$ is the fastest and contains more intrinsic noise than $Z$. In this regime, the mutual information between the components at the end of the cascade $I(X; Z)$ is larger than the mutual information between the abundance of either pair of directly connected components $I(X; Y),I(Y;Z)$.

How an additional ``read-out'' step can increase mutual information between molecular abundances can be intuitively understood in the regime in which the input signal varies on slower timescales than the lifetime of downstream cellular components, i.e., $\tau_x \gg \tau_y, \tau_z$ as is expected for many biological input signals. Under the above Gaussian approximation, the mutual information between $X$ and $Y$ is then given by
\begin{equation}
\label{EQ: XY Mutual Information Tx Large}
I(X; Y) = \frac{1}{2}\log_2 \left( 1 + \frac{\eta_\mathrm{sig}  }{\eta_{y}^\text{int}}\right)\hspace{3pt}.
\end{equation}
In contrast, the mutual information between $X$ and $Z$ is given by
\begin{equation}
\label{EQ: XZ Mutual Information Tx Large Limit}
\begin{split}
I(X; Z) &= \frac{1}{2}\log_2 \left( 1 + \frac{\eta_\mathrm{sig}  }{\eta_{z}^\text{int}}
\frac{1}{1 + \mfrac{\eta_{y}^\text{int} }{\eta_{z}^\text{int}} \mfrac{\tau_y}{\tau_y +  \tau_z}} \right)\hspace{3pt}.
\end{split}
\end{equation}
We see the magnitude of $I(X; Y)$ is limited by $\eta_\mathrm{sig} /\eta_{y}^\text{int}$, while the magnitude of $I(X; Z)$ is  limited by $\eta_\mathrm{sig} /\eta_{z}^\text{int}$ when $Z$ becomes slow.
The additional read-out step thus performs time-averaging that effectively replaces the intrinsic noise of $Y$ with that of $Z$. How much time-averaging is needed to violate Eq.~\ref{EQ: DPI XY XZ} depends on the ratio of intrinsic noises as given by Eq.~\ref{EQ: Mutual Information Condition}.

This suggests the increase in mutual information across more distant variables in a cascade occurs because the final slow variable filters out fast intrinsic noise of intermediates as previously reported for stochastic biochemical reaction cascades \cite{Paulsson2004,Paulsson2005,Taniguchi2010,Bokes2012}
Such time-averaging is common in biology and has, e.g., been reported in bacterial sensing \cite{berg1977physics}. \rcomm{It is equivalent to low-pass filters used in signal processing to remove high-frequency noise in engineering applications \cite{haykin2007signals}.}

\subsection*{Optimal timescale for maximizing mutual information}
The above results show that adding a component to a biochemical cascade can increase the mutual information between the stationary state abundances of the last component and the upstream input. Next, we determine the optimal timescale over which the additional variable $Z$ should average out fluctuations to maximize $I(X; Z)$.

For distributions that are approximately Gaussian, we find for the ideal read-out variable with $\eta_z^\text{int}=0$ the optimal timescale $\tau_z^*$ is given by
\begin{equation}
\begin{split}
\label{EQ: Optimal Tz}
\tau_z^* =\tau_x &\sqrt{\frac{\mfrac{\tau_y}{\tau_x} -2\left(\mfrac{\tau_y}{\tau_x}\right)^3  - \left(\mfrac{\tau_y}{\tau_x}\right)^2 \left(1 + 2 \mfrac{\eta_\mathrm{sig} }{\eta_{y}^\text{int}}\right)}{\left(1 + \mfrac{\tau_y}{\tau_x}\right)\left(\mfrac{\eta_\mathrm{sig} }{\eta_{y}^\text{int}} + \mfrac{\tau_y}{\tau_x} \right)}}
\end{split} \hspace{3pt} .
\end{equation}

The existence of this optimum can be intuitively understood because in the limit of $\tau_z \to \infty$, $Z$ averages out both intrinsic fluctuations from the intermediate as well as the signal, see Fig.~\ref{FIG:XYXZ Timescales}B. %
For read-out variables with finite noise, the optimal timescale $\tau_z^*$ monotonically decreases as $\eta_{z}^\text{int}$ increases, see Fig.~\ref{FIG:XYXZ Timescales}C, until ultimately violations of Eq.~\ref{EQ: DPI XY XZ} are no longer possible for any value of $\tau_z$, see %
Appendix \ref{APP: Mutual Information Two Step Appendix}. %

\subsection*{On-off upstream}
\begin{figure*}[hbt!]
\vspace{-1.4em}
\centering
\includegraphics[width=1.0\textwidth]{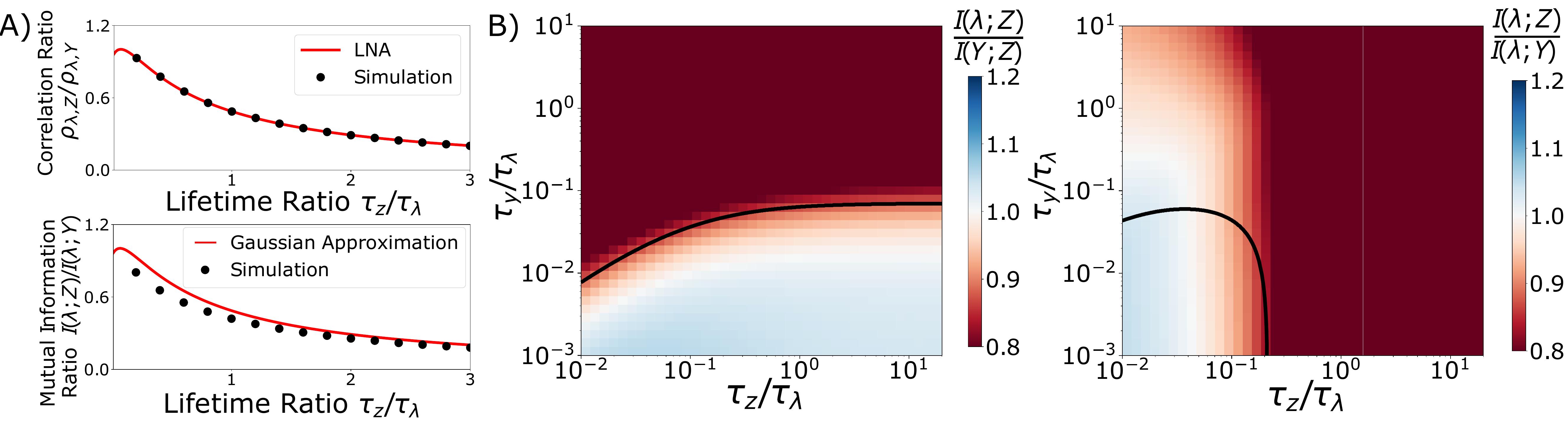}
\vspace{-1.4em}
\caption{\textbf{Gaussian approximations make qualitatively correct but quantitatively inaccurate predictions for mutual information in cascades responding to on-off signals.} %
A) Comparing exact numerical simulations with analytical results for the cascade defined by Eq.~\ref{EQ: definition of stochastic signal}. Correlation coefficients are exactly predicted by the linear noise approximation (LNA) because all rates are linear. Mutual information is only approximately predicted by Gaussian approximations. Data corresponds to a cascade with abundances $ \langle y \rangle = 2, \langle z \rangle = 20$ and $\tau_y / \tau_\lambda = 0.1,P_\text{on} = 0.5$.
B) Exact numerical simulations show the Gaussian approximation (black line) no longer quantitatively predicts the region of violations, but qualitative timescale features are accurately characterized, i.e., for noisy intermediates Eq.~\ref{EQ: DPI YZ XZ} is not obeyed when the intermediate $Y$ is significantly faster than both $X, Z$, and Eq.~\ref{EQ: DPI XY XZ} is not obeyed when the signal timescales are longer lived than all other components. Same parameters as panel (A). 
}
\label{FIG:Switch Upstream}
\end{figure*}
Instead of a Poissonian input signal, we next consider a two-state input signal, e.g., motivated by a single receptor that is in a bound/unbound state, or by a gene stochastically switching between two transcriptional states. We thus analyze the following cascade in which the production rate of $Y$ stochastically switches on and off according to
\begin{equation}
\begin{gathered}
\label{EQ: definition of stochastic signal}
\lambda =0  \myxrightleftharpoons{k_\mathrm{off}}{k_\mathrm{on}}  \lambda = 1
\\
\begin{array}{rcl}
y\hspace{-3pt} & \myrightarrow{\alpha\hspace{.1em} \lambda}& \hspace{-3pt}y + 1 \\
y\hspace{-3pt} & \myrightarrow{ y /\tau_y}& \hspace{-3pt}y - 1
\end{array}\hspace{8pt}
\begin{array}{rcl}
z\hspace{-3pt}& \myrightarrow{\beta y} & \hspace{-3pt}z + 1 \\
z\hspace{-3pt} & \myrightarrow{z /\tau_z} & \hspace{-3pt}z - 1 
\end{array}\hspace{3pt}.
\end{gathered}
\end{equation}

Exact numerical simulations of the process defined in Eq.~\ref{EQ: definition of stochastic signal} show that Eqs.~\ref{EQ: DPI YZ XZ}, \ref{EQ: DPI XY XZ} are also violated in this strongly non-Gaussian biochemical cascade, see Fig.~\ref{FIG:Switch Upstream}B.
Note, the stochastic switching process of Eq.~\ref{EQ: definition of stochastic signal} yields identical mathematical forms for the covariances and correlation coefficients as the original model given by Eq.~\ref{EQ: Two Step Cascade} with the only difference that variability of the input signal is now given by
\begin{equation}
\label{EQ: Intrinsic Noise With Switch Upstream}
\eta_{\lambda} = \frac{1-P_\text{on}}{P_\text{on} }\hspace{3pt},
\end{equation}
where $P_\text{on}$ is the fraction of time the gene spends in the on-state, and the signal timescale is given by \mbox{$\tau_\lambda =  1/(k_{\text{on}} + k_{\text{off}})$}. As far as correlation coefficients are concerned, the systems of Eq.~\ref{EQ: Two Step Cascade} and Eq.~\ref{EQ: definition of stochastic signal} behave identically and can be exactly solved analytically using the standard linear noise approach \cite{Paulsson2004}. It is the Gaussian approximation for the mutual information that loses accuracy, see Fig.~\ref{FIG:Switch Upstream}A.

While the behavior is now quantitatively different from the simple analytical approximation, the qualitative behavior remains. Reversal of Eq.~\ref{EQ: DPI YZ XZ} requires an intermediate component that is significantly faster than both $X, Z$, and reversal of Eq.~\ref{EQ: DPI XY XZ} requires a cascades in which the signal timescale is longer than the lifetime of both $Y,Z$.

Differences between the Gaussian approximation and the exact simulation data also become apparent for the cascade define system Eq.~\ref{EQ: Two Step Cascade} when components are present in abundances lower than two molecules on average \cite{SI}. 

\subsection*{Other types of biochemical cascades}
In the previous sections we showed that mutual information can increase along the cascade defined in Eq.~\ref{EQ: Two Step Cascade}.
Next, we analyze mutual information along more complex cascades to analyze the effect of molecular conversion events, or proofreading steps. (For a generalization of Eq.~\ref{EQ: Two Step Cascade} to four components, see supplement \cite{SI}.)

\subsubsection{Cascades with molecular conversion events}
\begin{figure}[!hbt]
\vspace{-1.em}
\centering
\includegraphics[width=0.95\columnwidth]{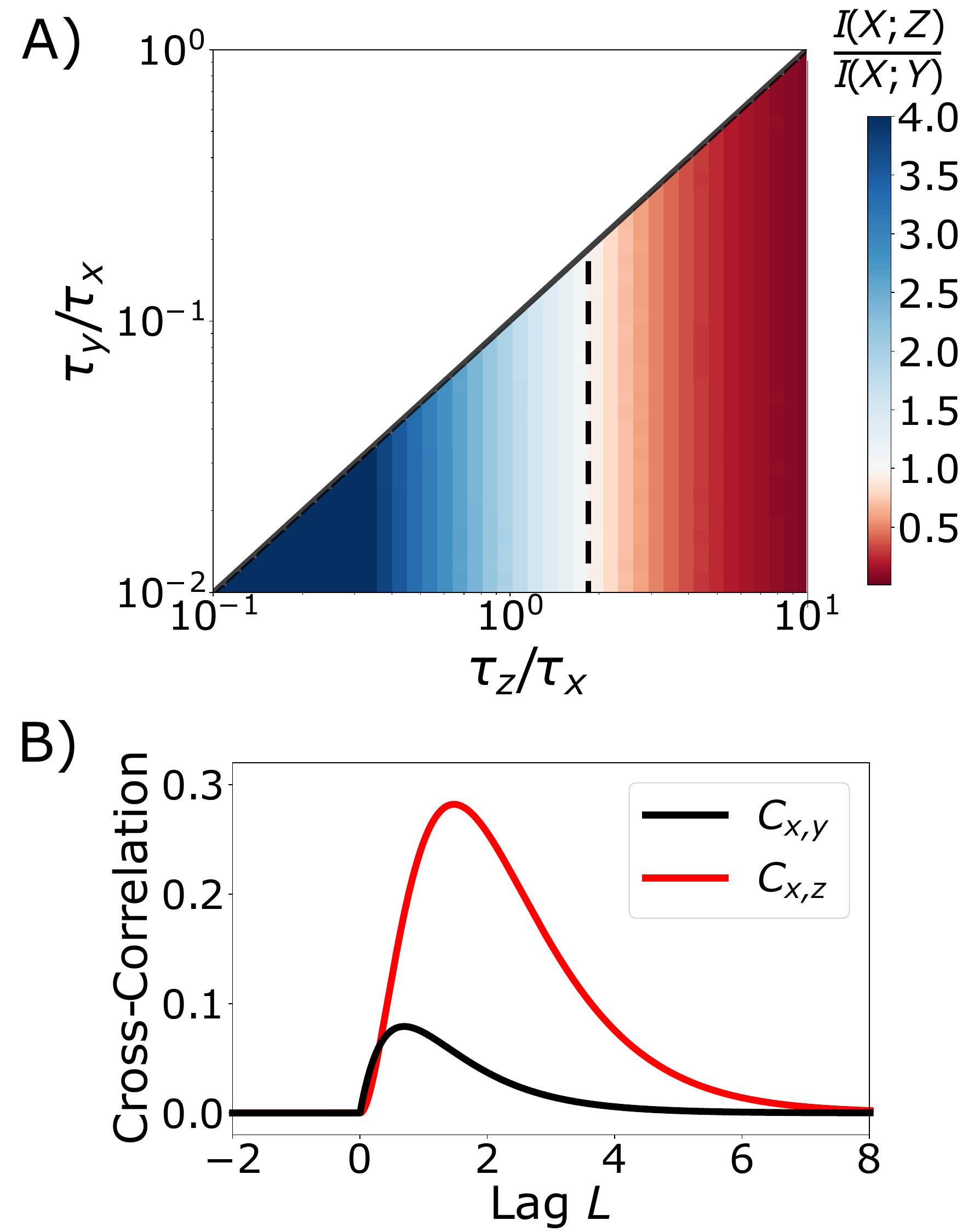}
\caption{\textbf{Cascades with conversion events also exhibit mutual information reversal that depends on relative timescales.} A) Exact numerical simulation results for the model defined by Eq.~\ref{EQ: Two Step Cascade Conversion} where $Y$-molecules are born through catalytic production and subsequently convert into $Z$-molecules. Data correspond to cascade with $\langle x \rangle = \langle z \rangle = 20$, and $\langle y \rangle = 2$. Cascades with conversion reactions and fixed averages are constrained in molecular lifetimes, resulting in the triangular accessible region. The (almost) vertical black dashed line corresponds to the Gaussian prediction for 
$I(X; Z) = I(X; Y)$ which closely matches the numerical data.
B) Temporal cross-correlations between components in a linear cascade in which $X$ converts into $Y$ which converts into $Z$. Lag is measured in units of $\tau_x$. At zero lag both cross-correlations are exactly zero because the steady state distributions are statistically independent. The causal connection between components only becomes apparent in temporal correlations with non-zero lags. Molecular abundances are the same as in panel (A) with lifetimes $\tau_x = 1, \tau_y = 0.5, \tau_z = 0.8$.
}
\label{FIG:Conversion Model}
\end{figure}

In the cascade defined by Eq.~\ref{EQ: Two Step Cascade} the levels of one molecule set the production rate of another. However, in biochemical cascades one molecule might convert into another through conformational changes.
We first consider the case where the intermediate variable $Y$ converts into the final read-out variable $Z$ by replacing the production reaction of $Z$ and the degradation reaction of $Y$ in Eq.~\ref{EQ: Two Step Cascade} with
\begin{equation}
\label{EQ: Two Step Cascade Conversion}
\begin{array}{rcl}
(y, z)\hspace{-3pt} & \myrightarrow{\beta_1 \hspace{.1em}y}& \hspace{-3pt}(y - 1, z + 1)
\\y\hspace{-3pt} & \myrightarrow{\beta_2 \hspace{.1em}y}& \hspace{-3pt}y - 1
\end{array}\hspace{3pt} ,
\end{equation}
where the lifetime of the intermediate component is now given by $\tau_y = 1/\left( \beta_1 + \beta_2 \right)$.

Exact numerical simulations of this cascade show that Eq.~\ref{EQ: DPI XY XZ} is now violated over a larger range of timescales, see Fig.~\ref{FIG:Conversion Model}A.
This occurs because $Z$ no longer inherits intrinsic noise from $Y$, see Appendix \ref{APP: Conversion Cascades}. This is most easily seen in the limit where $\tau_x \gg \tau_y, \tau_z$, where the Gaussian approximation now gives
\begin{equation}
\label{EQ: XZ Mutual Information Tx Large Limit, Conversion}
I(X; Z) =  \frac{1}{2}\log_2 \bigg( 1 + \frac{\eta_\mathrm{sig}  } {\eta_{z}^\text{int}}\bigg) \hspace{3pt}.
\end{equation}
Compared to Eq.~\ref{EQ: XZ Mutual Information Tx Large Limit}, which required $\tau_z \gg \tau_y$ for the read-out variable to remove the effect of intrinsic noise from the intermediate, in a cascade in which $Y$ converts into $Z$ the final read out removes the intermediate noise at all lifetimes. $I(X; Z)$ is always larger than in the case without conversions, which explains why violations occur over a wider range of timescales, see Appendix \ref{APP: Conversion Cascades}.

In the above cascade with a final conversion event, $I(X; Z)$ decreases monotonically with $\tau_z$. 
There is no advantage of time-averaging with larger $\tau_z$ when the intermediate noise is already filtered out via the conversion reaction, see Appendix \ref{APP: Conversion Cascades}.

Next, we consider the cascade in which $X$ converts into $Y$ but $Z$ is made catalytically. This corresponds to replacing the production reaction of $Y$  and degradation reaction of $X$ in Eq.~\ref{EQ: Two Step Cascade} with
\begin{equation}
\begin{array}{rcl}
(x, y)\hspace{-3pt} & \myrightarrow{\alpha_1 \hspace{.1em}x}& \hspace{-3pt}(x - 1, y + 1)
\\x\hspace{-3pt} & \myrightarrow{\alpha_2 \hspace{.1em}x}& \hspace{-3pt}x - 1
\end{array}\hspace{3pt} ,
\label{EQ: Conversion Event X Y}
\end{equation}
where the lifetime of the signal is now $\tau_x = 1/\left( \alpha_1 + \alpha_2 \right)$.

In this cascade the stationary state distribution factorizes with $P(x, y, z) = P(x) \times P(y, z)$ where $P(x)$ is a Poisson distribution, and $P(y, z)$ is the stationary state distribution of a system in which $Y$ is made at a constant rate reported previously \cite{li2021steady}, see Appendix \ref{APP: Conversion Cascades}. The initial conversion event thus makes the stationary state abundances of downstream components statistically independent of the signal and there is zero mutual information between the abundance of $X$ and its downstream variables for any parameter values.

To illustrate why the concept of statistical independence can be dramatically misleading when applied to stationary state distributions of stochastic processes we consider the case of a conversion cascade in which both $Y$ and $Z$ molecules are made in conversion events. This cascade belongs to a family of reaction networks for which the stationary state distribution is the product of three independent Poisson distributions \cite{anderson2010product, baez2012quantum}, which implies that the mutual information between any pair of variables is strictly zero \cite{tostevin2010mutual}. %

The physical dependence of these causally connected components is obscured by the statistical independence of their stationary state distributions.%
The interactions between biochemical components only become apparent in their temporal cross-correlation with lag $L$
\begin{equation}
\label{EQ: Cross Correlation Formula}
C_{i, j} (L) = \frac{\langle x_i (t) x_j (t + L) \rangle - \langle x_i (t) \rangle \langle x_j (t) \rangle}{\sqrt{\text{Var}(x_i) \text{Var}(x_j)}}\hspace{3pt},
\end{equation}
see Fig.~\ref{FIG:Conversion Model}B and Appendix \ref{APP: Conversion Cascades}. This highlights how statistical concepts developed for static random variables can be highly misleading when applied to stationary state distributions of dynamically varying components in biochemical reaction networks.

When considering the stationary-state distribution, an initial conversion event will seemingly disconnect the fluctuations of the input variable from any downstream variable regardless of the length of the cascade. This can be shown by considering  the general linear cascade $X_1 \rightarrow X_2 \rightarrow ... \rightarrow X_k$, where $X_1$ is produced at a constant rate, and all molecules undergo first order degradation reactions while arrows denote first order production rates that can be a molecular conversion reactions or simple one component setting the production rate of the next. For all such cascades the normalized covariance between components monotonically decreases over the cascade according to
\begin{equation}
\label{EQ: Covariance inequality}
\frac{\Cov(x_1, x_k)}{\langle x_1 \rangle \langle x_k \rangle} = \frac{\Cov(x_1, x_{k-1})}{\langle x_1 \rangle \langle x_{k - 1} \rangle} \frac{1}{1 + \mfrac{\tau_{k}}{\tau_{1}}} \hspace{3pt},
\end{equation}
where $\tau_1, \tau_k$ denote the average lifetimes of the first and $k^\mathrm{th}$ molecules in the chain. The above formula holds for $k \geq 3$, see Appendix \ref{APP: Conversion Cascades}. Because a cascade that starts with a conversion event has $\text{Cov}(x_1, x_2) = 0$, Eq.~\ref{EQ: Covariance inequality} implies that all downstream covariances vanish too.

In contrast, the \emph{rate} of mutual information transfer has been shown to be large through conversion events \cite{tostevin2010mutual}, highlighting how the mutual information of stationary state distributions cannot be used as a convenient proxy for information transfer in biochemical systems.

\subsubsection{Kinetic proofreading}
Finally, we consider kinetic proofreading \cite{Hopfield1974, ninio1975kinetic}, which can enhance the response of cells to different ligand occupancies and is commonly argued to allow cells to better distinguish between two ligands. However, recent work has shown that the increased average differences come at the cost of increased intrinsic noise which generally worsens the ability to distinguish between different ligands \cite{kirby2023proofreading}. 

\begin{figure}[!htb]
\centering
\includegraphics[width=1\columnwidth]{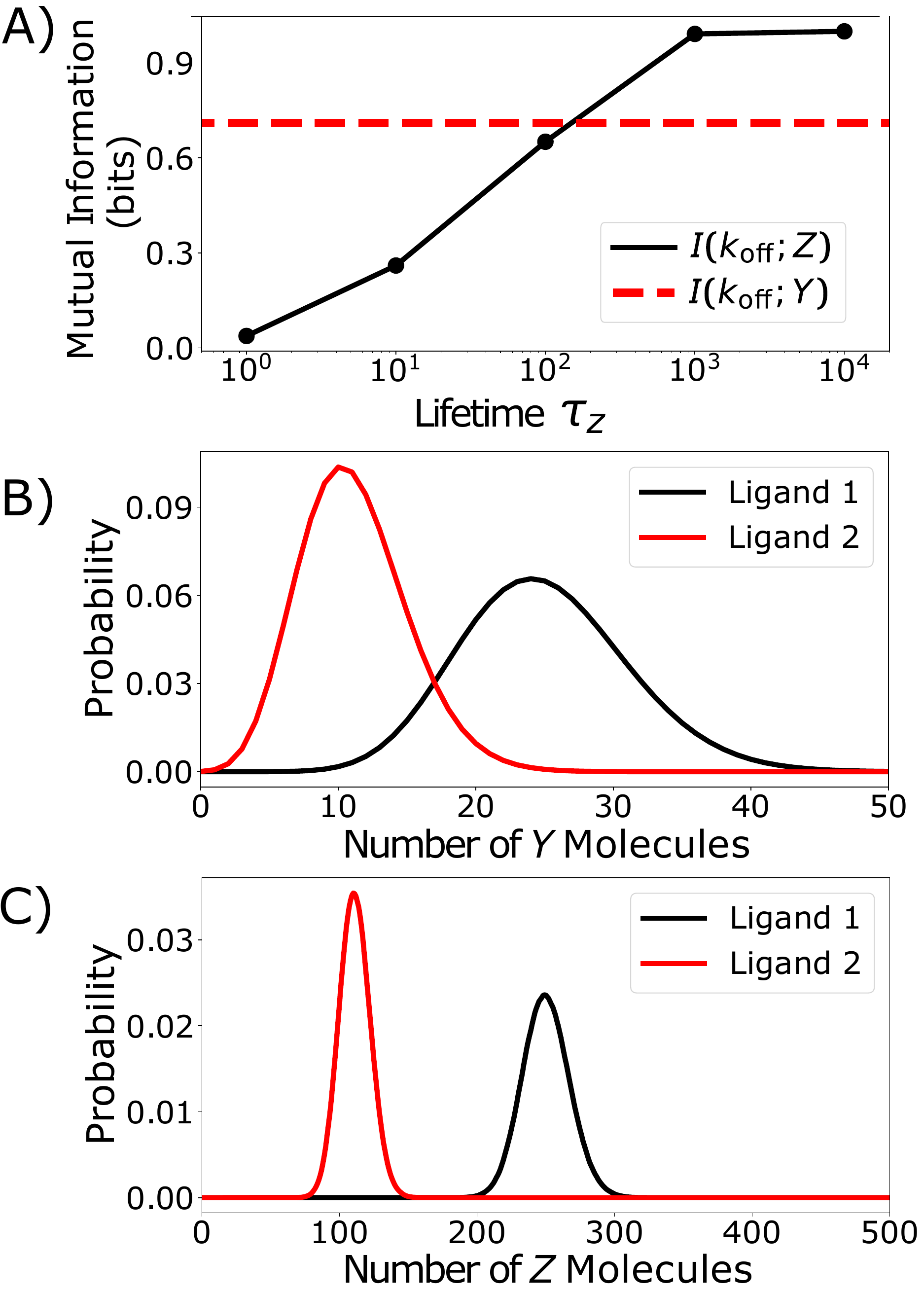}
\vspace{-1.4em}
\caption{\textbf{Time-averaging can alleviate the mutual information decreasing effect of kinetic proofreading.}
A) Simulation results for the kinetic proofreading model defined in the main text. Mutual information between the receptor output and the ligand dependent unbinding rate $k_\text{off}$ can be increased by adding an extra component $Z$ that time-averages the receptor output.
B) The probability distribution of the receptor output. Although different ligands result in different output averages, intrinsic noise causes significant overlap of their output distributions. C) Time-averaging effectively reduces the impact of intrinsic noise in $Y$ as illustrated by the more separated probability distribution of the final read-out variable $Z$.}
\label{FIG:KPF Figure}
\end{figure}

The effect of increased intrinsic noise in kinetic proofreading has been quantified through computing mutual information between ligand affinity and receptor output \cite{kirby2023proofreading}. We thus analyze the mutual information in the following model: ligands bind at a diffusion limited rate $k_{\text{on}}$, advance from the initial bound state to subsequent proofreading states at a rate $k_f$, and unbind from any state with a ligand dependent rate $k_{\text{off}}$. In the final bound state the receptor produces downstream signaling molecules $Y$ at a rate $k_p$. The unbinding rate $k_{\text{off}}$ depends on the ligand affinity while all other are assumed to be ligand independent. Considering two different ligands we consider the mutual information between downstream components and the unbinding rate $k_{\text{off}}$ that takes two different values with equal probability on a timescale much slower than $\tau_y,\tau_z$.

The results of the previous sections suggest that the effect of increased intrinsic noise in proofreading can be removed through time-averaging with an additional reaction step $Y \rightarrow Z$, as long as $\tau_z \gg \tau_y$. 
Indeed, numerical simulation results confirm this intuition for a single receptor with two potential ligands of different affinities such that $I(k_{\text{off}}, Z)$ is larger than $ I(k_{\text{off}}, Y)$, see Fig.~\ref{FIG:KPF Figure}A.

This result can be intuitively understood by looking at the probability distribution of output molecules. Although proofreading enhances the differences in average output associated with each ligand type, intrinsic noise can cause the resulting output distributions to overlap \cite{kirby2023proofreading}, see Fig.~\ref{FIG:KPF Figure}B. Adding an additional processing step with a slow read-out component can effectively reduce the impact of intrinsic noise through time-averaging, see Fig.~\ref{FIG:KPF Figure}C.
Intrinsic noise can also be reduced by increasing the lifetime of the read-out molecule directly instead of adding an additional processing step. Both methods of increasing the mutual information come at the expense of slowing down the system response to changes in ligand concentration highlighting a general speed-accuracy trade-off in sensing systems as remarked previously \cite{lan2012energy}.

\section*{Discussion}
\rcomm{The data processing inequality plays a crucial role in information theory because it sets a hard bound on the rate of information transfer between any two components that use a particular channel as an intermediary. %
}
In contrast, the mutual information between stationary state distributions of cellular abundances is not a measure of information flow \cite{tostevin2010mutual} but a statistical measure of similarity between the distributions of molecular levels which can be increased by post-processing. In special cases it can be shown that this measure obeys strict inequalities \cite{SI}, but in contrast to the fundamental data processing inequality such constraints are system specific.

Here, we clarify under which conditions the mutual information between stationary state distributions of molecular abundances behaves non-monotonically along biochemical signaling cascades.
\rcomm{By numerically determining stationary state probability distributions of biochemical cascades we characterize the timescale dependence of the non-monotonic behavior of mutual information between molecular abundances.}
We complement the exact numerical data with approximate analytical expressions
by approximating the steady state distribution as multivariate Gaussian distribution similar to previous work \cite{tostevin2010mutual, Biswas2016, Nandi2018, Nandi2019, Roy2021}. Direct comparison with numerical simulations shows that the \rcomm{qualitative behavior} of mutual information between components in biochemical cascades is well described by such approximations even for \rcomm{strongly non-Gaussian} cascades.

\rcomm{We find that the mutual information between stationary state distributions can increase along signaling cascades with fast and noisy intermediates 
as long as the upstream signal is sufficiently slow.
This is precisely the biologically relevant regime because for many cellular processes, the timescale of the signal is expected to be much longer than the lifetimes of the intracellular components.}
\rcomm{The mutual information between stationary state distributions of molecular abundances thus seems of limited utility for characterizing signaling processes in biology despite its convenient accessibility from experimental single-cell data.} 

\section*{Author Contributions}
RF derived the analytical results and performed the numerical simulations. RF and AH wrote the article.

\section*{Declaration of Interests}
The authors declare that they do not have any competing interests.

\section*{Acknowledgments}
We thank B Kell, Seshu Iyengar, Euan Joly-Smith for many helpful discussions and suggestions. %
This work was supported by the Natural Sciences and Engineering Research Council of Canada and a New Researcher Award from the University of Toronto Connaught Fund. 

\section*{Appendix}
\renewcommand{\thesubsection}{\Alph{subsection}}

\subsection{Two-step cascade correlations and mutual information}
\label{APP: Two Step Cascade Correlations and Mutual Information}
For chemical reaction systems with linear rates all moments can be exactly derived from the chemical master equation. In particular, the matrix of normalized (co)variances $\eta_{ij} := \Cov(x_i,x_j)/(\lb x_i \rb \lb x_j\rb)$ satisfies the Lyapunov equation \cite{van1992stochastic,Paulsson2004,Paulsson2005,hilfinger2015a}
\begin{equation}
M\eta+(M\eta)^T + D=0 \hspace{3pt} ,
\label{EQ: Lyapunov General}
\end{equation}
where $M$ and $D$ for the stochastic reaction system defined by Eq.~\ref{EQ: Two Step Cascade} are given by
\begin{align}
\label{EQ: Definition of M and D for two step cascade}
M\hspace{-2pt} &= \hspace{-2pt}\left(
\begin{array}{ccc}
 -\dfrac{1}{\tau _x} & 0 & 0 \\
\dfrac{1}{\tau _y} & -\dfrac{1}{\tau _y} & 0 \\
0 & \dfrac{1}{\tau _z} & -\dfrac{1}{\tau _z} \\
\end{array}
\right) 
\\D &= \hspace{-2pt}
\left(
\begin{array}{ccc}
 \dfrac{2}{\tau _x \langle x\rangle } & 0 & 0 \\
0 & \dfrac{2}{\tau _y \langle y\rangle } & 0 \\
 0 & 0 & \dfrac{2}{\tau _z \langle z\rangle } \\
\end{array}
\right)\hspace{3pt}.
\nonumber
\end{align}

Using Eqs.~(\ref{EQ: Lyapunov General},\ref{EQ: Definition of M and D for two step cascade}), one can solve for the following normalized (co)variances
\begin{equation}
\begin{split}
\label{EQ: Two Step Covariances}
& \eta_{xx} = \frac{1}{\langle x \rangle}, \hspace{70pt} \eta_{xy} = \frac{1}{\langle x \rangle}\frac{1}{1 + \mfrac{\tau_y}{\tau_x}}
\\& \eta_{yy} = \frac{1}{\langle y \rangle} + \frac{1}{\langle x \rangle}\frac{1}{1 + \mfrac{\tau_y}{\tau_x}}, \quad \eta_{xz} = \frac{1}{\langle x \rangle}\frac{1}{1 + \mfrac{\tau_y}{\tau_x}}\frac{1}{1 + \mfrac{\tau_z}{\tau_x}}
\\& \eta_{yz} = \frac{1}{\langle y \rangle}\frac{1}{1 + \mfrac{\tau_z}{\tau_y}} + \frac{1}{\langle x \rangle}\frac{1}{1 + \mfrac{\tau_y}{\tau_x}} \left(\frac{1}{1 + \mfrac{\tau_z}{\tau_x}}\frac{1}{1 + \mfrac{\tau_y}{\tau_z}} + \frac{1}{1 + \mfrac{\tau_z}{\tau_y}}\right)
\\& \text{and}
\\& \eta_{zz} = \frac{1}{\langle z \rangle} + \eta_{yz}\hspace{3pt}.
\end{split}
\end{equation}

The correlation coefficients can in turn be obtained from
\begin{equation}
\label{EQ: Correlations From Normalized Covariances}
\rho_{xy} = \frac{\eta_{xy}}{\sqrt{\eta_{xx} \eta_{yy}}} \hspace{3pt},
\end{equation}
which yields
\begin{equation}
\begin{split}
\label{EQ: Two Step Correlation Coefficients}
\rho_{xy} = &\frac{1}{1 + \mfrac{\tau_y}{\tau_x}} \Biggl(\frac{1}{\mfrac{\langle x \rangle}{\langle y \rangle} + \mfrac{\tau_x}{\tau_x + \tau_y}}\Biggr)^{1/2}
\\\rho_{xz} = &\Bigg[\frac{1}{1 + \mfrac{\tau_y}{\tau_x}}\bigg(\frac{1}{1 + \mfrac{\tau_z}{\tau_y}} + \frac{1}{1 + \mfrac{\tau_y}{\tau_z}} \frac{1}{1 + \mfrac{\tau_z}{\tau_x}}\bigg)
\\& + \frac{\langle x \rangle}{\langle z \rangle} + \frac{\langle x \rangle}{\langle y \rangle}\frac{1}{1 + \mfrac{\tau_z}{\tau_y}}\Bigg]^{-1/2} \hspace{-3pt}\times \frac{1}{1 + \mfrac{\tau_y}{\tau_x}}\frac{1}{1 + \mfrac{\tau_z}{\tau_x}}
\\\rho_{yz} = & \Bigg[\frac{\langle x \rangle}{\langle y \rangle}\frac{1}{1 + \mfrac{\tau_z}{\tau_y}} + \frac{1}{1 + \mfrac{\tau_y}{\tau_x}} \Bigg(\frac{1}{1 + \mfrac{\tau_z}{\tau_x}}\frac{1}{1 + \mfrac{\tau_y}{\tau_z}} + \frac{1}{1 + \mfrac{\tau_z}{\tau_y}}\Bigg)\Bigg]
\\& \times \Bigg[\Bigg( \frac{1}{1 + \mfrac{\tau_y}{\tau_x}} \bigg(\frac{1}{1 + \mfrac{\tau_z}{\tau_x}}\frac{1}{1 + \mfrac{\tau_y}{\tau_z}} + \frac{1}{1 + \mfrac{\tau_z}{\tau_y}}\bigg)
\\& + \frac{\langle x \rangle}{\langle z \rangle} + \frac{\langle x \rangle}{\langle y \rangle}\frac{1}{1 + \mfrac{\tau_z}{\tau_y}}\Bigg)\bigg(\frac{\langle x \rangle}{\langle y \rangle} + \frac{1}{1 + \mfrac{\tau_y}{\tau_x}} \bigg)\Bigg]^{-1/2} .
\end{split}
\end{equation}

\subsubsection{Mutual Information}
\label{APP: Mutual Information Two Step Appendix}
Approximating the stationary state distribution of Eq.~\ref{EQ: Two Step Cascade} as a multivariate Gaussian with the same second order moments allows us to obtain approximate analytic expressions for the mutual information via
\begin{equation}
\label{EQ: Gaussian Mutual Information Correlation Coefficient Relation}
I(X; Y) = - \frac{1}{2}\log_2 \bigg(1 - \rho_{xy}^2\bigg)\hspace{3pt} .
\end{equation}

The expressions obtained from applying Eq.~\ref{EQ: Gaussian Mutual Information Correlation Coefficient Relation} to the correlation coefficients of Eq.~\ref{EQ: Two Step Correlation Coefficients} lead to the inequalities Eq.~\ref{EQ: XZ YZ Correlation Inequality} and Eq.~\ref{EQ: DPI XY XZ - Simplified Correlations} via elementary algebraic manipulations. 

We find this approximation accurately describes the mutual information in our cascade unless average abundances fall below two molecules at which point the discreteness of the distribution becomes apparent \cite{SI}.

To show that the lifetime ratio $\tau_z^* / \tau_x$ that maximizes the mutual information $I(X; Z)$ is a decreasing function of $\eta_{z}^\text{int}/\eta_\mathrm{sig} $, we note that the mutual information given by Eq.~\ref{EQ: Gaussian Mutual Information Correlation Coefficient Relation} is an increasing function of the correlation coefficient. 
When $\eta_{z}^\text{int}/\eta_\mathrm{sig}  \rightarrow 0$, the correlation coefficient is given by
\begin{equation}
\begin{split}
\rho_{xz} \hspace{-.5ex}=\hspace{-.5ex} &\left(\frac{1}{1 + \mfrac{\tau_y}{\tau_x}}\hspace{-.1ex}\Bigg(\frac{1}{1 + \mfrac{\tau_z}{\tau_y}} \hspace{-.5ex}+\hspace{-.5ex} \frac{1}{1 + \mfrac{\tau_y}{\tau_z}} \frac{1}{1 + \mfrac{\tau_z}{\tau_x}}\Bigg) \hspace{-.5ex}+\hspace{-.5ex} 
\frac{\eta_{y}^\text{int}}{\eta_\mathrm{sig}}
\frac{1}{1 + \mfrac{\tau_z}{\tau_y}}\right)^{\hspace{-.3em}-1/2} 
\\& \times \frac{1}{1 + \mfrac{\tau_y}{\tau_x}}\frac{1}{1 + \mfrac{\tau_z}{\tau_x}} .
\end{split}
\end{equation}
for which a nonzero optimal time scale $\tau_z^*$ (given by Eq.~\ref{EQ: Optimal Tz}) can exist because the denominator can decrease faster with $\tau_z$ than the numerator due to the $\eta_{y}^\text{int}/\eta_\mathrm{sig}$ term. 

In the regime $\eta_{z}^\text{int}/\eta_\mathrm{sig}  \gg \eta_{y}^\text{int}/\eta_\mathrm{sig} , 1$ we have
\begin{equation}
\begin{split}
\rho_{xz} = & \left(\frac{\eta_\mathrm{sig} }{\eta_{z}^\text{int}}\right)^{1/2} \times \frac{1}{1 + \mfrac{\tau_y}{\tau_x}}\frac{1}{1 + \mfrac{\tau_z}{\tau_x}}
\end{split}
\end{equation}
for which the optimal time scale is $\tau_z^* = 0$, as the correlation coefficient is a decreasing function of $\tau_z$. Increasing $\eta_{z}^\text{int}/\eta_\mathrm{sig} $ reduces the effect of the $\eta_{y}^\text{int}/\eta_\mathrm{sig}$ term monotonically, resulting in smaller optimal $\tau_z^*$ until eventually $\tau_z^* = 0$ .

\subsubsection{Mutual Information Inequalities}
\label{APP: Mutual Information Inequalities}
In the regime $\tau_z, \tau_y \ll \tau_x $, Eq.~\ref{EQ: XZ YZ Correlation Inequality} becomes
\begin{equation}
\begin{split}
1 + \frac{\tau_y}{\tau_z}\left(1 + \frac{\eta_\mathrm{sig} }{\eta_{y}^\text{int}}\right) \leq \left( 1 + \frac{\tau_y}{\tau_z} \right) \left(1 + \frac{\eta_\mathrm{sig} }{\eta_{y}^\text{int}} \right)^{1/2}
\end{split}\hspace{3pt},
\end{equation}
from which Eq.~\ref{EQ: XZ YZ Diagonal} follows from algebraic manipulations. 

Equality of Eq.~\ref{EQ: XZ YZ Correlation Inequality} implies
\begin{equation}
\label{EQ: Ty as a Function of Tz, Inequality 1}
\begin{split}
\frac{\tau_z}{\tau_x} = \frac{-\mfrac{\tau_y}{\tau_x} \left(1 + \mfrac{\eta_\mathrm{sig} }{\eta_{y}^\text{int}} \left(1 + \mfrac{\tau_y}{\tau_x} \right) - \sqrt{\mfrac{\eta_\mathrm{sig} }{\eta_{y}^\text{int}} + \left(1 + \mfrac{\tau_y}{\tau_x}\right)^{-1}}\right)}{1 + \mfrac{\tau_y}{\tau_x}\left(1 + \mfrac{\eta_\mathrm{sig} }{\eta_{y}^\text{int}}\left(1 + \mfrac{\tau_y}{\tau_x} \right)\right) - \sqrt{\mfrac{\eta_\mathrm{sig} }{\eta_{y}^\text{int}} + \left(1 + \mfrac{\tau_y}{\tau_x}\right)^{-1}}}
\end{split}\hspace{3pt}.
\end{equation}
When $\tau_z \gg \tau_x$, Eq.~\ref{EQ: Ty as a Function of Tz, Inequality 1} implies that 
\begin{equation}
1 + \frac{\tau_y}{\tau_x}\left(1 + \frac{\eta_\mathrm{sig} }{\eta_{y}^\text{int}}\left(1 + \frac{\tau_y}{\tau_x} \right)\right) - \sqrt{\frac{\eta_\mathrm{sig} }{\eta_{y}^\text{int}} + \left(1 + \frac{\tau_y}{\tau_x}\right)^{-1}} = 0 .
\end{equation}
which is a fifth order polynomial in $\tau_y / \tau_x$ whose roots are analytically intractable.

However, we can obtain the roots in $\eta_\mathrm{sig}  / \eta_{y}^\text{int}$, given by
\begin{equation}
\frac{\eta_\mathrm{sig} }{\eta_{y}^\text{int}} = \frac{1 - 2 \mfrac{\tau_y}{\tau_x} - 4\mfrac{\tau_y}{\tau_x}^2 - 2\mfrac{\tau_y}{\tau_x}^3 \pm \sqrt{1 - 4\mfrac{\tau_y}{\tau_x} - \mfrac{\tau_y}{\tau_x}^2}}{2 \left(\mfrac{\tau_y}{\tau_x}^2 + 2\mfrac{\tau_y}{\tau_x}^3 + \mfrac{\tau_y}{\tau_x}^4 \right)} \hspace{3pt}.
\end{equation}
Positivity of the variability ratios then gives rise to the necessary condition of \mbox{$\tau_y / \tau_x \leq 1/2 (\sqrt{2} - 1)$} to obtain equality in Eq.~\ref{EQ: XZ YZ Correlation Inequality}.

To obtain Eq.~\ref{EQ: Noise Limit Condition}, we note the terms in Eq.~\ref{EQ: DPI XY XZ - Simplified Correlations} that are not proportional to variability measures can be grouped into a positive term on the left. Removing them yields the following necessary inequality
\begin{equation}
\frac{\eta_{z}^\text{int}}{\eta_\mathrm{sig} }\left(1 + \frac{\tau_z}{\tau_x}\right)^2 
 \leq \frac{\eta_{y}^\text{int}}{\eta_\mathrm{sig}} \left(1 - \frac{\left(1 + \mfrac{\tau_z}{\tau_x}\right)^2}{1 + \mfrac{\tau_z}{\tau_y}}\right)\hspace{3pt}.
\end{equation}
The right side becomes larger without the term in the brackets, and algebraic manipulations then yield Eq.~\ref{EQ: Noise Limit Condition}.

In the limit $\tau_z \ll \tau_x$, Eq.~\ref{EQ: DPI XY XZ - Simplified Correlations} becomes
\begin{equation}
 1 + \frac{\tau_z}{\tau_y} 
 \leq \frac{\eta_{y}^\text{int}}{\eta_{z}^\text{int}}\frac{\tau_z}{\tau_y} + \frac{\eta_\mathrm{sig} }{\eta_{z}^\text{int}}\frac{\tau_z}{\tau_y}\frac{1}{1 + \mfrac{\tau_y}{\tau_x}}\hspace{3pt}.
\end{equation}
Applying the additional limit $\tau_y \ll \tau_x$ results in Eq.~\ref{EQ: Mutual Information Condition}.

When $\eta_y^\text{int} \gg \eta_\mathrm{sig} ,\eta_z^\text{int}$ Eq.~\ref{EQ: DPI XY XZ - Simplified Correlations} simplifies to 
\begin{equation}
0 \leq 1 - \frac{\left(1 + \mfrac{\tau_z}{\tau_x}\right)^2}{1 + \mfrac{\tau_z}{\tau_y}} \quad ,
\end{equation}
from which Eq.~\ref{EQ: Simple Lifetime Condition} follows from algebraic manipulation.

\subsection{Cascades with conversion events}
\label{APP: Conversion Cascades}
Applying Eq.~\ref{EQ: Lyapunov General} to the model defined in Eq.~\ref{EQ: Two Step Cascade} and Eq.~\ref{EQ: Two Step Cascade Conversion} yields the following normalized covariances
\begin{equation}
\begin{split}
\label{EQ: Two Step Conversion Covariances}
& \eta_{xx} = \frac{1}{\langle x \rangle}, \hspace{70pt} \eta_{xy} = \frac{1}{\langle x \rangle}\frac{1}{1 + \frac{\tau_y}{\tau_x}}
\\& \eta_{yy} = \frac{1}{\langle y \rangle} + \frac{1}{\langle x \rangle}\frac{1}{1 + \frac{\tau_y}{\tau_x}}, \quad \eta_{xz} = \frac{1}{\langle x \rangle}\frac{1}{1 + \frac{\tau_y}{\tau_x}}\frac{1}{1 + \frac{\tau_z}{\tau_x}}
\\& \eta_{yz} = \frac{1}{\langle x \rangle}\frac{1}{1 + \frac{\tau_y}{\tau_x}} \left(\frac{1}{1 + \frac{\tau_z}{\tau_x}}\frac{1}{1 + \frac{\tau_y}{\tau_z}} + \frac{1}{1 + \frac{\tau_z}{\tau_y}}\right)
\\& \eta_{zz} = \frac{1}{\langle z \rangle} + \frac{1}{\langle x \rangle}\frac{1}{1 + \mfrac{\tau_y}{\tau_x}} \left(\frac{1}{1 + \mfrac{\tau_z}{\tau_x}}\frac{1}{1 + \mfrac{\tau_y}{\tau_z}} + \frac{1}{1 + \mfrac{\tau_z}{\tau_y}}\right)
\end{split} .
\end{equation}
Computing the correlation coefficients from these covariances then straightforwardly yields approximate expressions for the mutual information.

The resulting approximate $I(X; Z)$ monotonically decreases with $\tau_z$, because Eq.~\ref{EQ: Gaussian Mutual Information Correlation Coefficient Relation} which is an increasing function of the correlation coefficient given by
\begin{align}
\rho_{xz} = &\left(\frac{1}{1 + \mfrac{\tau_y}{\tau_x}}\left(\frac{1}{1 + \mfrac{\tau_z}{\tau_y}} + \frac{1}{1 + \mfrac{\tau_y}{\tau_z}} \frac{1}{1 + \mfrac{\tau_z}{\tau_x}}\right) + \frac{\eta_{z}^\text{int}}{\eta_\mathrm{sig}}\right)^{\hspace{-.2em}-1/2} \nonumber
\\& \times \frac{1}{1 + \mfrac{\tau_y}{\tau_x}}\frac{1}{1 + \mfrac{\tau_z}{\tau_x}} .
\end{align}
which in turn is a decreasing function of $\tau_z$.

For the model defined in Eq.~\ref{EQ: Two Step Cascade} and Eq.~\ref{EQ: Conversion Event X Y}, the chemical master equation is given by
\begin{equation}
\begin{split}
\label{EQ: XY Converts, Chemical Master Equation}
\frac{dP(x, y, z)}{dt} &= - \Bigg[\lambda + \frac{x}{\tau_x} + 
y\left(\beta+\frac{1}{\tau_y}\right) + \frac{z}{\tau_z}\Bigg]P(x, y, z)
\\& \hspace{-4ex}+ \lambda P(x-1, y, z) + \alpha_2 \left(x + 1 \right) P(x+1, y, z)
\\& \hspace{-4ex}+ \alpha_1 (x + 1) P(x+1, y-1, z) + \frac{y + 1}{\tau_y} P(x, y+1, z)
\\& \hspace{-4ex}+ \beta y P(x, y, z-1) + \frac{z + 1}{\tau_z} P(x, y, z+1) \hspace{3pt}.
\end{split}
\end{equation}
Substituting the ansatz $P(x, y, z) = P(x) \times P(y, z)$ into the stationary state condition of the above master equation with  the Poissonian
\begin{equation}
P(x) = \frac{\langle x \rangle^x e^{-\langle x \rangle}}{x!} \hspace{3pt},
\end{equation}
results in the following condition
\begin{equation}
\begin{split}
0= &-\Bigg[\alpha_1 \langle x \rangle + y\left( \beta + \frac{1}{\tau_y} \right) + \frac{z}{\tau_z} \Bigg]P(y, z) 
\\& + \alpha_1 \langle x \rangle P(y-1, z) + \frac{y + 1}{\tau_y} P(y+1, z) 
\\& + \beta y P(y, z-1) + \frac{z+1}{\tau_z} P(y, z+1)
\end{split}\hspace{3pt}.
\label{EQ: Factorized CME for conversion cascade}
\end{equation}
which is satisfied as long as
$P(y, z)$ is the stationary state distribution that solves the chemical master equation of a system
in which $Y$ is made at a constant rate $\alpha_1 \langle x \rangle$ and linearly affects the production of $Z$ molecules with a rate $\beta y$. 
This implies that $P(x, y, z) = P(x) \times P(y, z)$ is indeed the stationary state distribution of the model defined in Eq.~\ref{EQ: Two Step Cascade} and Eq.~\ref{EQ: Conversion Event X Y} as claimed in the main text.

The cross-correlations of the conversion cascade in which both $Y$ and $Z$ molecules are made in conversion events can be derived from the corresponding chemical master equation \cite{Heuett2006}. For $L<0$ the temporal cross-correlations are zero and for $L\geq0$ we obtain
\begin{equation}
\begin{split}
C_{x, y} (L) &= %
\sqrt{\frac{\eta_\mathrm{sig}}{\eta_{y}^\text{int}}}\frac{e^{-L/\tau_y} - e^{-L/\tau_x}}{1 - \mfrac{\tau_x}{\tau_y}} 
\\C_{x, z}(L) &= \frac{1}{\tau_x \tau_y} 
\sqrt{\frac{\eta_\mathrm{sig}}{\eta_{z}^\text{int}}}
\left(\left(\frac{1}{\tau_x} - \frac{1}{\tau_y}\right)\left(\frac{1}{\tau_x} - \frac{1}{\tau_z}\right)\left(\frac{1}{\tau_y} - \frac{1}{\tau_z}\right) \right)^{-1}
\\& \times \bigg(e^{-L/\tau_z}\left(\frac{1}{\tau_x} - \frac{1}{\tau_y}\right) + e^{-L/\tau_x}\left(\frac{1}{\tau_y} - \frac{1}{\tau_z}\right)
\\& + e^{-L/\tau_y}\left(\frac{1}{\tau_z} - \frac{1}{\tau_x}\right)\bigg) \hspace{3pt},
\end{split}
\end{equation}
which are plotted in the main text. 

To derive Eq.~\ref{EQ: Covariance inequality}, we utilize the fluctuation balance equations that must be satisfied by any pairs of components $X_i$ and $X_j$ in a system with stationary probability distributions \cite{hilfinger2015a} 
\begin{equation}
\Cov(x_i, R_j^{-} - R_j^{+}) + \Cov(x_j, R_i^{-} - R_i^{+}) = \sum_{k = 1}^{N} s_{ki}s_{kj} \langle r_{k} \rangle ,
\label{EQ: Covariance Rate Equation}
\end{equation}
where $x_i$ is the abundance of molecule $X_i$, $R_{j}^{\pm}$ is the total flux of production or degradation for molecule $X_j$, and $\langle r_{k} \rangle$ is the average reaction rate for reaction $k$ in which levels of $X_i$ are changed by $s_{ki}$. 

For the linear cascade $X_1 \rightarrow X_2 \rightarrow ... \rightarrow X_k$, Eq.~\ref{EQ: Covariance Rate Equation} yields the following relation for $k > 2$ 
\begin{equation}
\Cov(x_1, -R_k) + \Cov(x_k, -R_1) = 0 ,
\end{equation}
with reaction fluxes are given by
\begin{equation}
R_1 = \lambda - \frac{x_1}{\tau_1}, R_k = c_k x_{k - 1} - \frac{x_k}{\tau_k} \hspace{3pt},
\end{equation}
where $\lambda$ is the production rate of the first molecule, $c_k x_{k - 1}$ is the rate at which molecule $X_{k - 1}$ converts into molecule $X_k$, and $\tau_k$ the lifetime of the $k^\mathrm{th}$ molecule. Algebraic manipulations then result in Eq.~\ref{EQ: Covariance inequality}.%

\bibliography{bibliography}
\end{document}


\date{}
\maketitle

In this document we present the results of numerical simulations of the system defined by Eq.~1 for abundances different than those reported in Fig.~1 and Fig.~3. Furthermore, we present the corresponding analysis of longer cascades with four components, and derive an example bound that constrains the mutual information between stationary state distributions in a simple cascade.

\newpage

\section*{Numerical simulations of of Eq.~1 For Different Abundances}
We simulate the model defined by Eq.~1 for several average abundances. First, we explore the effect of different $Z$ abundances compared to the main text.
\begin{figure*}[htbp!]
\centering
\includegraphics[width=0.84\linewidth]{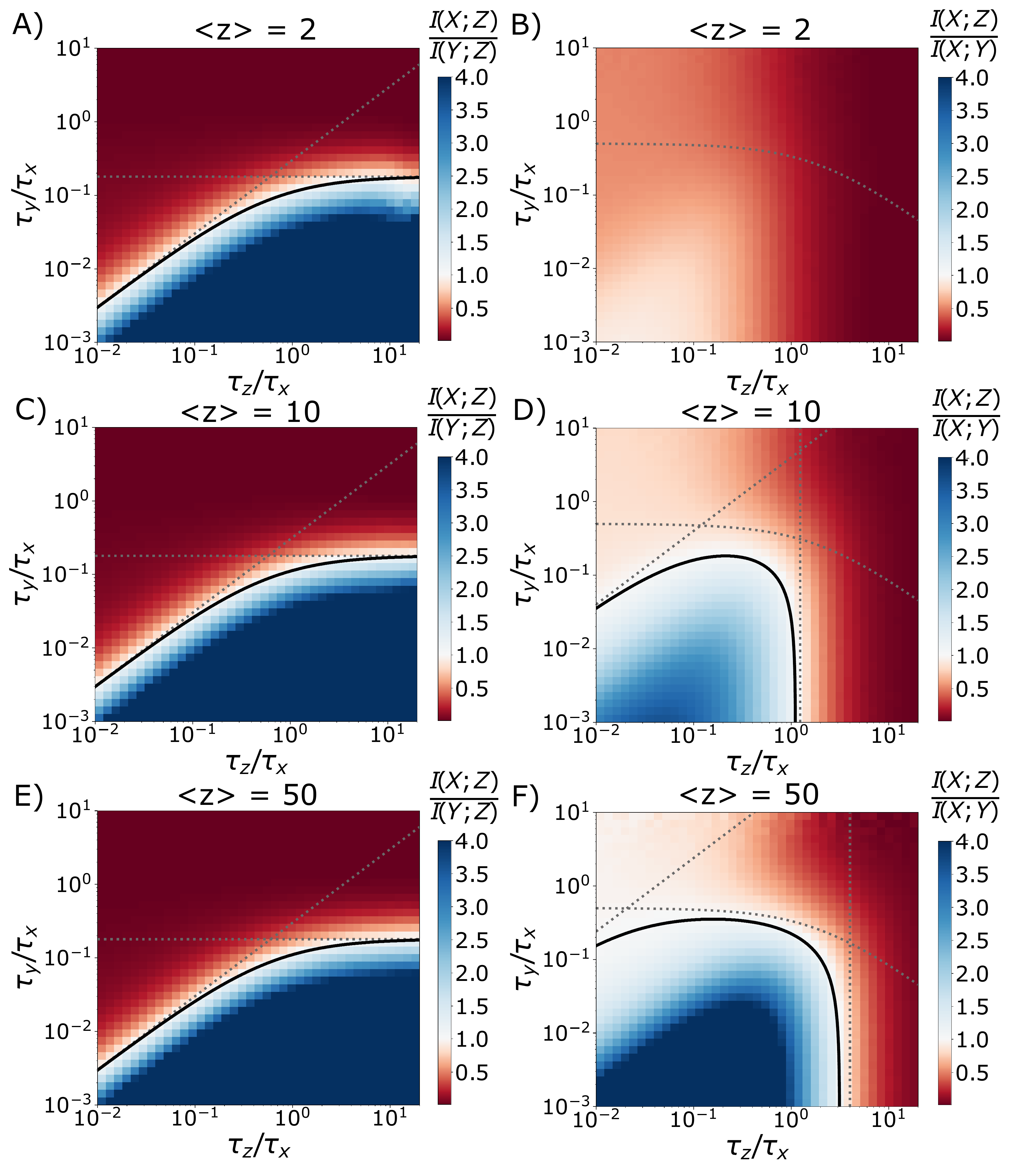}
\caption{\textbf{Simulating Eq.~1 for different $Z$ abundances.} We plot exact numerical simulation results for the mutual information between pairs of variables in the cascade defined by Eq.~1. Blue region indicates the parameter regime in which Eq.~2 and Eq.~3 of the main text are violated. Plots are generated for $\langle x \rangle = 20, \langle y \rangle = 2$ and varying $\langle z \rangle$. A) C) E) correspond to $I(X; Z)/I(Y; Z)$, where the solid black line corresponds to Eq.~4 of the main text, and the dashed gray lines correspond to the simpler inequalities derived in the main text. There is little dependence on $\langle z \rangle$ here. B) D) F) correspond to $I(X; Z)/I(X; Y)$, where the solid black line corresponds to Eq.~8 of the main text, and the dashed gray lines correspond to the simpler inequalities of Eqs.~9--11 in the main text.
}
\label{FIG:Numerical Verification of Protein Variances Diff Z}
\vspace{-.5em}
\end{figure*}

Next, we analyze the effect of varying all averages by a constant scale factor $U$.
\begin{figure*}[htbp!]
\centering
\includegraphics[width=0.84\linewidth]{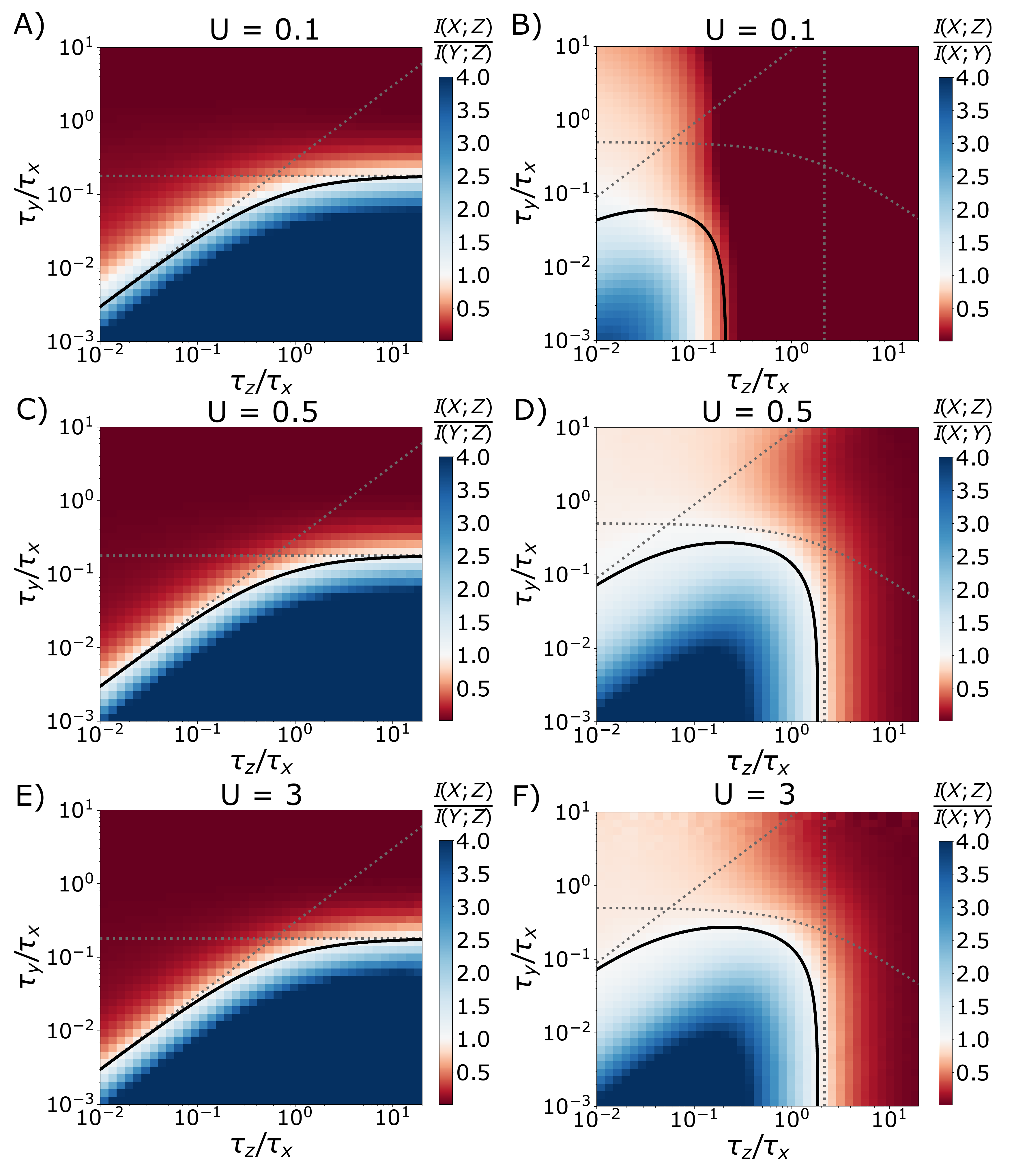}
\caption{\textbf{Simulating Eq.~1 for different abundance scales.} We plot exact numerical simulation results for the mutual information between pairs of variables in the cascade defined by Eq.~1. Blue region indicates the parameter regime in which Eq.~2 and Eq.~3 of the main text are violated. Plots are generated for $\langle x \rangle  = \langle z \rangle = 20U, \langle y \rangle = 2 U$, where $U$ is a scaling factor. A) C) E) correspond to $I(X; Z)/I(Y; Z)$, where the solid black line corresponds to Eq.~4 of the main text, and the dashed gray lines correspond to the simpler inequalities derived in the main text. We observe only a minor dependence on the behavior on the overall scale $U$. B) D) F) correspond to $I(X; Z)/I(X; Y)$, where the solid black line corresponds to Eq.~8 of the main text, and the dashed gray lines correspond to the simpler inequalities Eq.~9--11 in the main text.
}
\label{FIG:Numerical Verification of Protein Variances}
\vspace{-.5em}
\end{figure*}
\newpage
Finally, following our results for the stochastic switch model of Eq.~15, we simulate Eq.~1 with $\langle x \rangle = 0.5$, obtaining similar results to the plots of Fig.~4.

\begin{figure*}[hbt!]
\vspace{-1.4em}
\centering
\includegraphics[width=1.0\textwidth]{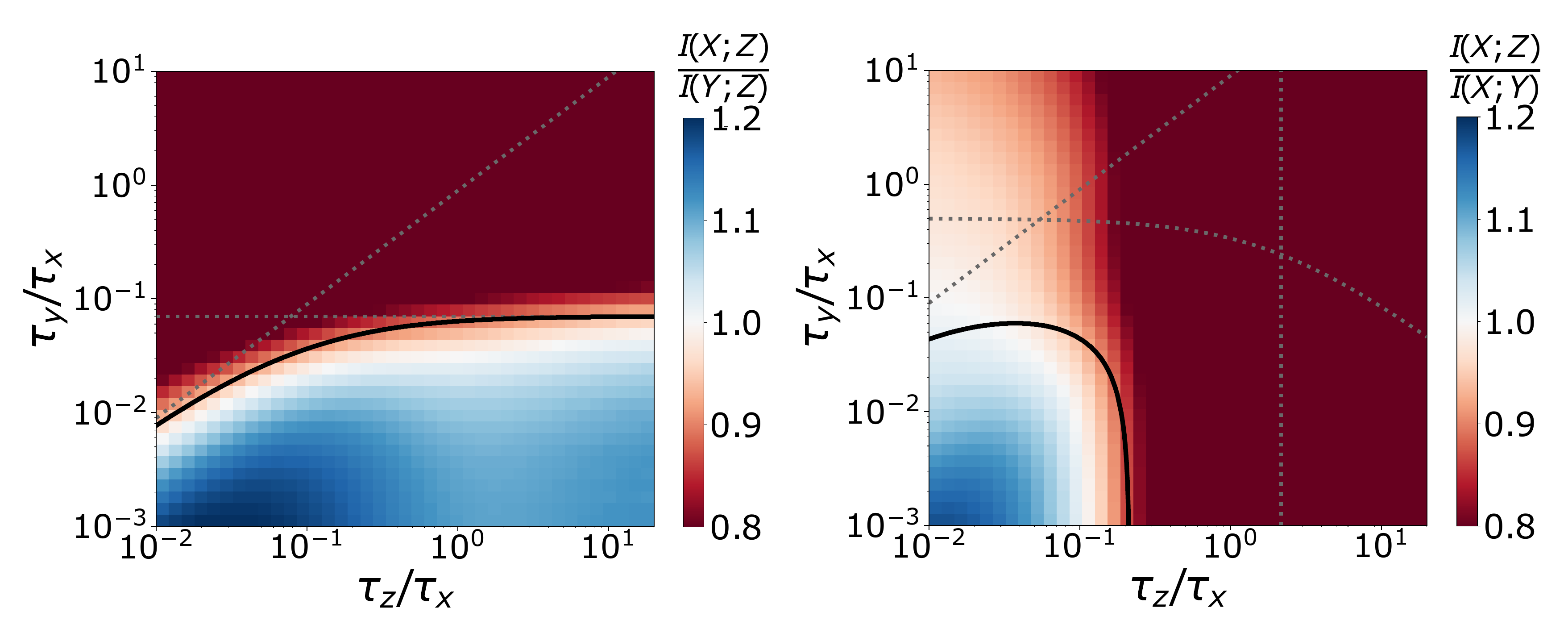}
\vspace{-1.4em}
\caption{\textbf{On-off upstream signals and Poissonian signals with low $X$ averages behave similarly. 
In both cases the Gaussian approximations make accurate qualitative but not quantitative predictions.}
B) Exact numerical simulations of the system defined by Eq.~1 in the main text for $\langle x \rangle = 0.5, \langle y \rangle = 2, \langle z \rangle = 20 $. The Gaussian approximation (black line) no longer quantitatively predicts the region of violations, but  qualitative time-scale features are accurately characterized, i.e., Eq.~2 is not obeyed when the intermediate $Y$ is significantly faster than both $X, Z$, and Eq.~3 is not obeyed when the signal timescales are longer lived than all other components. This is similar to the results obtained for the on-off upstream signal defined in the Eq.~15 of the main text and plotted in Fig.~4B.
}
\label{FIG:Switch Upstream}
\end{figure*}
\newpage
\section*{Three step cascades}

\begin{figure}[!b]
\vspace{-1.4em}
\centering
\includegraphics[width=1\columnwidth]{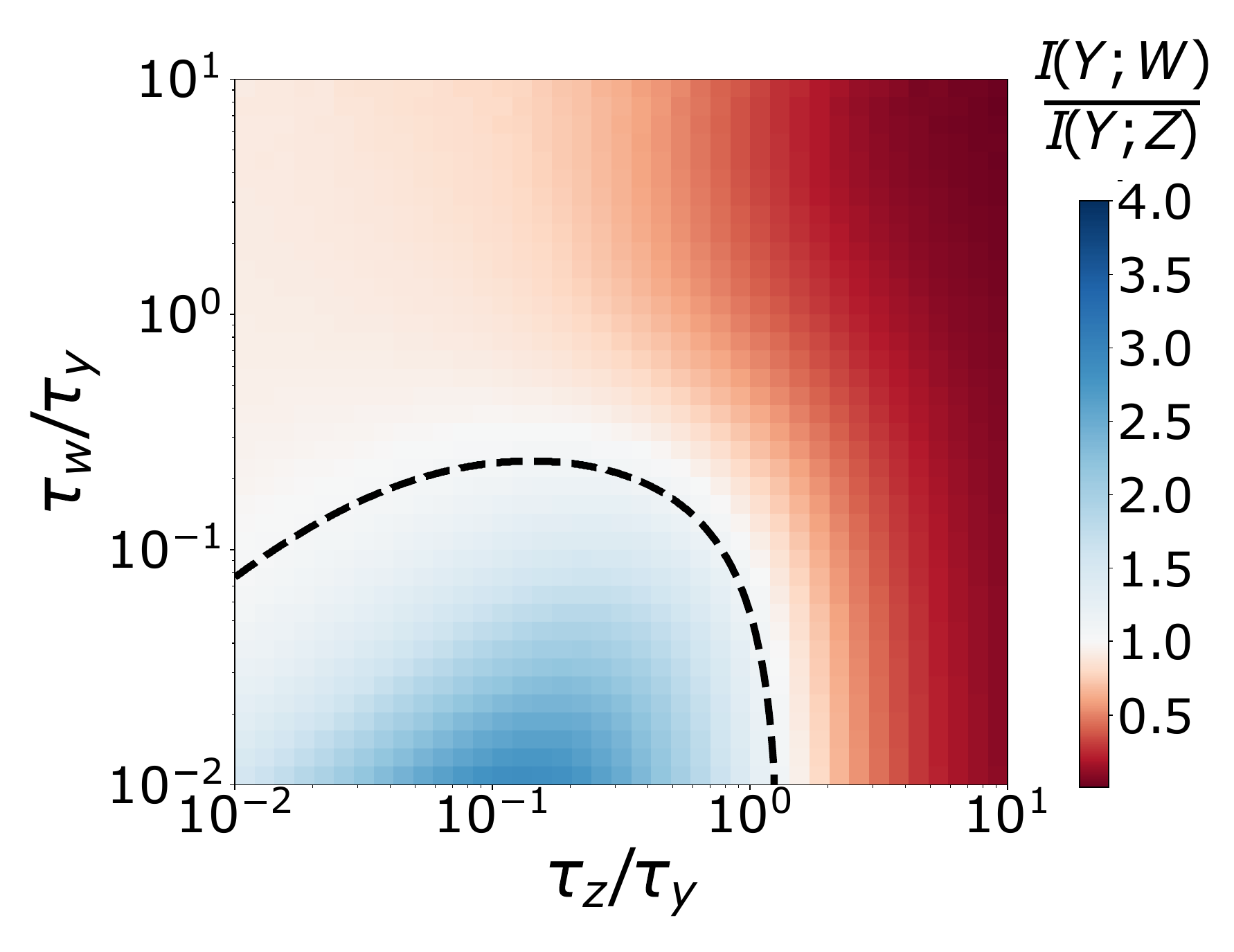}
\vspace{-1.4em}
\caption{\textbf{Non-monotonic behavior of mutual information in longer linear cascades}. Exact numerical simulation results for the ratio of mutual information $I(Y; W)/I(Y; Z)$ for the model defined in Eq.~1 of the main text and Eq.~S.1 of the supplement, reveals that $I(Y; W) > I(Y; Z)$ can occur. This shows that mutual information can increase along a cascade even when the upstream signal is no longer a Poissonian variable. The graph was generated for $\langle x \rangle = 0.5, \langle y \rangle = \langle w \rangle = 20, \langle z \rangle = 2, \tau_x = 0.1, \tau_y = 1$. 
}
\label{FIG:Larger Chains Model}
\end{figure}

Arguably, the analysis of the two-step cascade of Eq.~1 corresponds to a special case in which all tested inequalities involved the mutual information between a Poisson variable and a downstream component. We thus next analyze the mutual information between the second, third, and forth component in a three-step cascade. We consider the dynamics of Eq.~1 with the following additional reactions
\begin{equation}
\label{EQ: Three Step Cascade}
\begin{array}{rcl}
w\hspace{-3pt}& \myrightarrow{\gamma z} & \hspace{-3pt}w + 1 \\
w\hspace{-3pt} & \myrightarrow{w /\tau_w} & \hspace{-3pt}w - 1 
\end{array} \quad .
\end{equation}

Numerical simulations show that it is again possible for mutual information to be larger between more distant variables. The parameters for this regime are similar to the original two step cascade, where the intrinsic noise of the intermediate (now $Z$), and the lifetimes of downstream components must be small compared to the timescale of the upstream variability.

This cascade also allows us to study the behavior of mutual information over multiple intermediates, for which it has been previously shown that the mutual information can increase for certain parameters (Refs.~$[23,24]$ of the main text). Our analytical approximation suggests such violations occur because Eq.~13 generalizes to longer, meaning the noise of multiple intermediates can be eliminated depending on the relative lifetimes of the molecules involved. 

The covariances in the three step cascade are given by Eq.~25 along with the following additional covariances
\begin{equation}
\label{EQ: Three Strep Cascade Covariances}
\begin{split}
& \eta_{xw} = \frac{1}{\langle x \rangle}\frac{1}{1 + \mfrac{t_y}{t_x}}\frac{1}{1 + \mfrac{t_z}{t_x}}\frac{1}{1 + \mfrac{\tau_{w}}{\tau_x}}
\\& \eta_{yw} =\frac{1}{\langle x \rangle}\frac{1}{1 + \mfrac{\tau_y}{\tau_x}}\bigg(\frac{1}{1 + \mfrac{\tau_{w}}{\tau_y}} \bigg(\frac{1}{1 + \mfrac{\tau_z}{\tau_x}}\frac{1}{1 + \mfrac{\tau_y}{\tau_z}} + \frac{1}{1 + \mfrac{\tau_z}{\tau_y}}\bigg) + \frac{1}{1 + \mfrac{\tau_{y}}{\tau_{w}}}\frac{1}{1 + \mfrac{t_z}{t_x}}\frac{1}{1 + \mfrac{\tau_{w}}{\tau_x}}\bigg) + \frac{1}{\langle y \rangle}\frac{1}{1 + \mfrac{t_z}{t_y}}\frac{1}{1 + \mfrac{\tau_{w}}{\tau_y}}
\\& \eta_{zw} =\frac{1}{\langle x \rangle}\frac{1}{1 + \mfrac{\tau_y}{\tau_x}}\bigg(\frac{1}{1 + \mfrac{\tau_w}{\tau_z}}\bigg(\frac{1}{1 + \mfrac{\tau_z}{\tau_x}}\frac{1}{1 + \mfrac{\tau_y}{\tau_z}} + \frac{1}{1 + \mfrac{\tau_z}{\tau_y}}\bigg)  + \frac{1}{1 + \mfrac{\tau_z}{\tau_w}}\frac{1}{1 + \mfrac{\tau_{w}}{\tau_y}} \bigg(\frac{1}{1 + \mfrac{\tau_z}{\tau_x}}\frac{1}{1 + \mfrac{\tau_y}{\tau_z}} + \frac{1}{1 + \mfrac{\tau_z}{\tau_y}}\bigg)
\\&\hspace{30pt} + \frac{1}{1 + \mfrac{\tau_{y}}{\tau_{w}}}\frac{1}{1 + \mfrac{\tau_z}{\tau_x}}\frac{1}{1 + \mfrac{\tau_{w}}{\tau_x}}\bigg) + \frac{1}{\langle z \rangle}\frac{1}{1 + \mfrac{\tau_w}{\tau_z}} + \frac{1}{\langle y \rangle}\frac{1}{1 + \mfrac{\tau_z}{\tau_y}}\bigg(\frac{1}{1 + \mfrac{\tau_w}{\tau_z}} + \frac{1}{1 + \mfrac{\tau_{w}}{\tau_y}}\frac{1}{1 + \mfrac{\tau_{z}}{\tau_{w}}} \bigg)
\\& \eta_{ww} = \frac{1}{\langle w \rangle} + \eta_{zw} \quad .
\end{split}
\end{equation}
from which approximate expressions for the mutual information can be derived via the Gaussian approximation.

\section*{Bounds on mutual information between stationary state distributions}
\label{SEC: Static MI Bounds}

In this section, we derive and analyze a tight bound for the mutual information in the steady state distributions of all reaction networks with an on-off upstream connected via a linear production rate to arbitrarily complex downstream components, which is a generalization of the system defined in Eq.~15 of the main text.

A trivial bound on all mutual information is the entropy
\begin{equation}
\label{EQ: Mutual Information Definition Bound}
I(X; Y) \leq H(X), H(Y)
\end{equation}
which follows from the definition of mutual information, and does not depend on the details of the system like the data-processing inequality. Violations imply an error in calculating these quantities from the corresponding joint probability distribution, and are not helpful in analyzing biochemical systems.

Tighter bounds can be derived by specifying details of the system. When an upstream component $X$ is connected to downstream components through a linear production rate as follows

\begin{equation}
\label{EQ: Poisson Channel}
\begin{array}{rcl}
y\hspace{-3pt}&\myrightarrow{x}&\hspace{-3pt} y + 1 \\
\end{array}
\end{equation}
then the mutual information between $X$ and an arbitrary downstream component $Z$ that depends on $X$ only through $Y$ as an intermediary is bound by

\begin{equation}
\label{EQ: Birth Time Events Inequality}
I(X; Z) \leq I(X; \Vec{T} = (t_1, t_2, t_3...))
\end{equation}
where $\Vec{T}$ is an infinite dimensional vector that contains the times at which every $Y$ birth event is observed. This follows by definition as $Z$ can only obtain information about $X$ from the $Y$ birth events. For example, in the process defined by Eq.~1, $Z$ was a form of time-averaging of $Y$, where it effectively counts the more recent $Y$ birth events. However, time-averaging is one of many methods to construct a downstream variable $Z$, and does not necessarily maximize the mutual information.

While Eq.~\ref{EQ: Birth Time Events Inequality} is true for all networks with the reaction Eq.~\ref{EQ: Poisson Channel}, the distribution $P(X, \Vec{T})$ is experimentally inaccessible because it requires measuring an infinite number of $Y$ birth events. A finite number of event times can be obtained in a simulation with specified $X$ dynamics, but it is unclear how accurate finite representations of $P(X, T_{event})$ will affect the mutual information.

The problem of infinite dimensional distributions are avoided when considering an on-off signal of the form
\begin{equation}
\label{EQ: definition of on-off signal}
\begin{array}{rcl}
x_\text{off}  \myxrightleftharpoons{k_\mathrm{off}}{k_\mathrm{on}}  x_\text{on} 
\end{array}\hspace{3pt},
\end{equation}
for which the state of the upstream is known exactly when a birth event is observed. This form of signal was considered in Eq.~15 of the main text. For this signal, a birth event at $t = t_0$ implies $x(t_0) = x_{on}$, and there is no additional information on the future values of $X$ gained by knowing the history of $x$ beyond knowing the current value of $X$. This implies all information about the current value of $x$ is contained only in the time of most recent birth event, and thus $I(x; \Vec{T}) = I(X; T_{\text{last event}})$. The probability distribution $P(X, T_{\text{last event}})$ can be obtained from a Gillespie simulation, or directly from the chemical master equation.

Note it is not sufficient for only the current value of $X$ to be known exactly. Even a signal $X$ that can only take on two values can in general depend on a larger network. The history of $X$ can thus contain information about the state of this larger network, and $I(x; \Vec{T}) \neq I(x; T_{\text{last event}})$. For example, if $X$ is periodically driven, then past values of $X$ can be used to infer the period and phase.

\begin{figure}[!b]
\vspace{-1.4em}
\centering
\includegraphics[width=1\columnwidth]{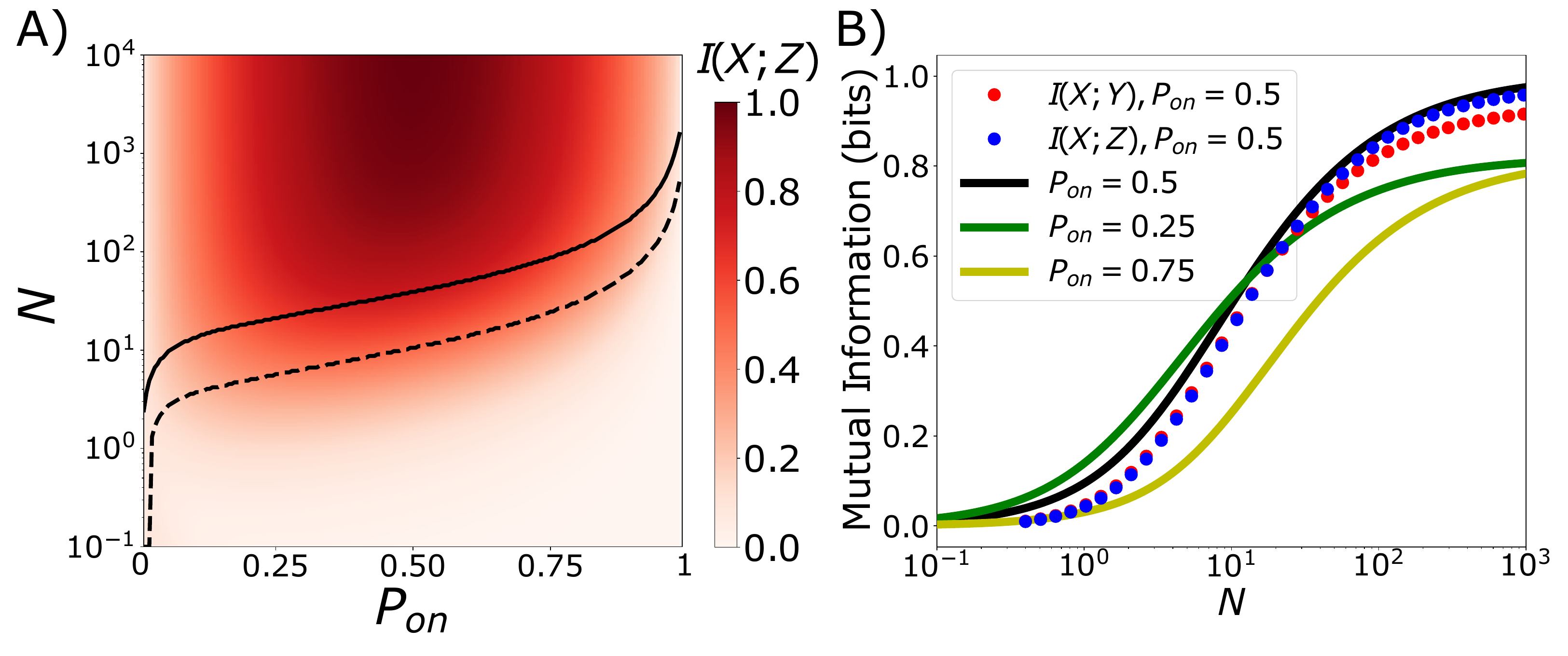}
\vspace{-1.4em}
\caption{\textbf{The maximum mutual information between a stochastic on-off signal $X$ and any downstream variable $Z$ depends on probability of the signal to be in its on-state $P_\text{on}$ and the average number of $Y$ molecules produced during the on-state of the signal, N.} A) For the upstream signal defined by Eq.~\ref{EQ: definition of on-off signal},  the maximum mutual information between the steady state distributions of $Z$ and any downstream variable $Z$ that depends on $X$ only through birth events of an intermediate molecule $Y$ that occur at rate $\alpha x$ is given by Eq.~\ref{EQ: Maximum Mutual Information}. Although this integral cannot be solved analytically, it can be plotted as a function of $\langle x \rangle = P_\text{on}$ and $N$. This plot shows that the mutual information rises with $N$ as expected, since more events provide more information for $Z$. The solid black line indicates where $I(X;Z) = 0.5 \times H(P_\text{on})$ while the dashed line indicates $I(X;Z) = 0.25 \times H(P_\text{on})$. $H(P_\text{on})$ is the binary entropy function, which is the maximum possible mutual information for this system. There is an asymmetry where smaller $P_\text{on}$ result in larger potential mutual information.
B) Plotting the maximum mutual information $I(X; Z)$ as a function of $N$ for several values of $P_\text{on}$ (colored lines) reveals that the mutual information rapidly increases for $N \geq 10$. We compare our theoretical prediction to results from simulating Eq.~\ref{EQ: definition of on-off signal} for $P_\text{on} = 0.5$ (dots), and we find the maximum mutual information achieved by time-averaging in our two step cascade is close to the theoretical limit when $N > 10$, but performs significantly worse when $N$ is small.}
\label{FIG: On-Off Switch Mutual Information}
\end{figure}

Proceeding with the analysis of the signal defined by Eq.~\ref{EQ: definition of on-off signal}, yields the following upper bound 
\begin{equation}
\begin{split}
\label{EQ: Maximum Mutual Information}
I(X;& Z) \leq H(P_\text{on}) + \frac{N P_\text{on}}{\sqrt{\big(N + \frac{1}{P_\text{off}}\big)^2 - 4 N \frac{P_\text{on}}{P_\text{off}}}} \times \int_{0}^{\infty} \bigg[ \big(e^{r_2 t} - e^{r_1 t}\big) \log \big( e^{r_2 t} -  e^{r_1 t}\big)
\\& + \bigg( \bigg(\frac{P_\text{on}}{P_\text{off}} + r_2 \bigg) e^{r_2 t} - \bigg(\frac{P_\text{on}}{P_\text{off}} + r_1 \bigg) e^{r_1 t}\bigg) \times \log \bigg(\big(\frac{P_\text{on}}{P_\text{off}} + r_2 \big) e^{r_2 t} - \big(\frac{P_\text{on}}{P_\text{off}} + r_1 \big) e^{r_1 t} \bigg)
\\&- \bigg(\big(\frac{1}{P_\text{off}} + r_2 \big) e^{r_2 t} - \big(\frac{1}{P_\text{off}} + r_1 \big) e^{r_1 t}\bigg) \times \log \bigg(\big(\frac{1}{P_\text{off}} + r_2 \big) e^{r_2 t} - \big(\frac{1}{P_\text{off}} + r_1 \big) e^{r_1 t}\bigg) \bigg] dt 
\end{split}
\end{equation}
where $P_\text{on}$ is the fraction of the time $x$ spends in the on state, $N$ is the average number of birth events that occur during one instance of the on state, $H(X)$ is the binary entropy function and $r_2, r_1$ are given by
\begin{equation}
r_{2, 1} = \frac{1}{2}\bigg(- N - \frac{1}{P_\text{off}} \pm \sqrt{\bigg(N + \frac{1}{P_\text{off}}\bigg)^2 - 4 N \frac{P_\text{on}}{P_\text{off}}}\bigg)
\end{equation}
The right-hand-side of Eq.~\ref{EQ: Maximum Mutual Information} cannot be solved analytically but it can be evaluated numerically as a function of $P_\text{on}$ and $N$, see Fig.~\ref{FIG: On-Off Switch Mutual Information}.

Intuitively, mutual information is an increasing function of $N$. A large rate of events implies that when events have not been observed for a long time (relative to characteristic timescale $\tau_x$) then the upstream must be in the off state. Conversely, a low event frequency $N$ makes it difficult to infer the current value of $x$. This intuition agrees with our solution as shown by Fig.~\ref{FIG: On-Off Switch Mutual Information}. Furthermore, for $N = 10000$ it can be shown that $I(x; z) \approx H(P_\text{on})$, see Appendix \ref{APP: Deriving Bounds on Mutual Information Between Static Distributions}.

The behavior at intermediate $N$ is more complicated. Although $I(X; Z)$ for very small and large $N$ are symmetric in $P_\text{on}$, at intermediate values an asymmetry occurs where mutual information $I(X; Z)$ is larger for $P_\text{on} \leq 0.5$ than for $P_\text{on} \geq 0.5$. As a result, although the maximum entropy occurs at $H(P_\text{on} = 0.5)$, the maximum mutual information for intermediate $N$ occurs at a lower value of $P_\text{on}$.

Since $N = \langle y \rangle \tau_y / \langle x \rangle \tau_x$ when $Y$ undergoes first order degradation, this bound can be directly applied to our previous results for the process defined in Eq.~15 of the main text, which had the same form of upstream signal as the model solved here. This comparison reveals that time-averaging performs fairly comparable to the theoretical limit for large $N$, but that there exist better methods of generating a $Z$ with high mutual information for low $N$. 

Although the bound of Eq.~\ref{EQ: Maximum Mutual Information} can be formulated in terms of experimentally measurable variables such as averages and lifetimes, it only applies to a specific linear cascade with an on-off input signal unlike the data-processing inequality which holds for all Markov Chains without any assumptions about their specific dependencies and input signals. 

Furthermore, the utility of the data-processing inequality is also not limited to computing bounds and ruling out particular network motifs. Bounding the mutual information across any network which uses a particular channel as an intermediary motivates the study of small network motifs as they yield an upper limit on larger networks. Meanwhile the approach outlined here only finds bounds on the mutual information between a specified signal and the variables downstream and does not apply if the signal of interest lies further upstream. 

Finally, note that the mutual information limited by the data-processing inequality is the information transferred across the network, which is easy to interpret and has clear relevance to biological systems. The mutual information between stationary state distributions considered here is less clear, as the steady state distributions are generated through sampling the actual temporal process through which communication occurs.

\subsection*{Derivation of Eq.~\ref{EQ: Maximum Mutual Information}}
\label{APP: Deriving Bounds on Mutual Information Between Static Distributions}

For the following model
\begin{equation}
\label{EQ: definition of stochastic signal - inequality}
\begin{array}{rcl}
x_\text{off}  \myxrightleftharpoons{k_\mathrm{off}}{k_\mathrm{on}}  x_\text{on} 
\end{array}\hspace{3pt},
\begin{array}{rcl}
y\hspace{-3pt} & \myrightarrow{\alpha\hspace{.1em} x_{on}}& \hspace{-3pt}y + 1
\end{array}\hspace{3pt}
\end{equation}
we derive the probability distribution $P(X, T_{\text{last event}})$. This is the probability of a particular $X$ state and time since the last event $t = T_{\text{last event}}$, and obeys the following ODEs
\begin{equation}
\begin{split}
\frac{dP(x_\text{on}, t)}{dt} &= - (\alpha + k_\text{off}) P(x_\text{on}, t) + k_\text{on} P(x_\text{off}, t)
\\\frac{dP(x_\text{off}, t)}{dt} &= - k_\text{on} P(x_\text{off}, t) + k_\text{off} P(x_\text{on}, t)
\end{split}
\end{equation}
where the two $X$ states flip between each other with rates $k_\text{on}, k_\text{off}$, and the total probability decreases through the $x_\text{on}$ state due to the possibility of a new birth event. The general solution is
\begin{equation}
P(1, t) = c_1 \frac{k_\text{on} + r_1}{k_\text{off}} e^{r_1 t} + c_2 \frac{k_\text{on} + r_2}{k_\text{off}} e^{r_2 t} 
\end{equation}
\begin{equation}
P(0, t) = c_1 e^{r_1 t} + c_2 e^{r_2 t} 
\end{equation}
where $c_1, c_2$ are undetermined constants, and the eigenvalues are 
\begin{equation}
r_{1, 2} = \frac{- \alpha - k_\text{off} - k_\text{on} \pm \sqrt{\alpha^2 + 2( k_\text{off} + k_\text{on} ) \alpha + k_\text{off}^2 + 2 k_\text{on} k_\text{off} + k_\text{on}^2}}{2}
\end{equation}

Note that for our system $P(0, t) = 0$ birth events can only occur if $X$ is in its on-state and thus $c_1 = -c_2$ Furthermore, the total probability of all possible $X$ states and times must sum to zero, which implies
\begin{equation}
1 = \int_{0}^{\infty} \big( P(0, t) + P(1, t) \big) dt = c_1 \left(\frac{k_\text{on} + k_\text{off}}{r_1 r_2} \right) .
\end{equation}

\begin{figure}[!b]
\vspace{-1.4em}
\centering
\includegraphics[width=1\columnwidth]{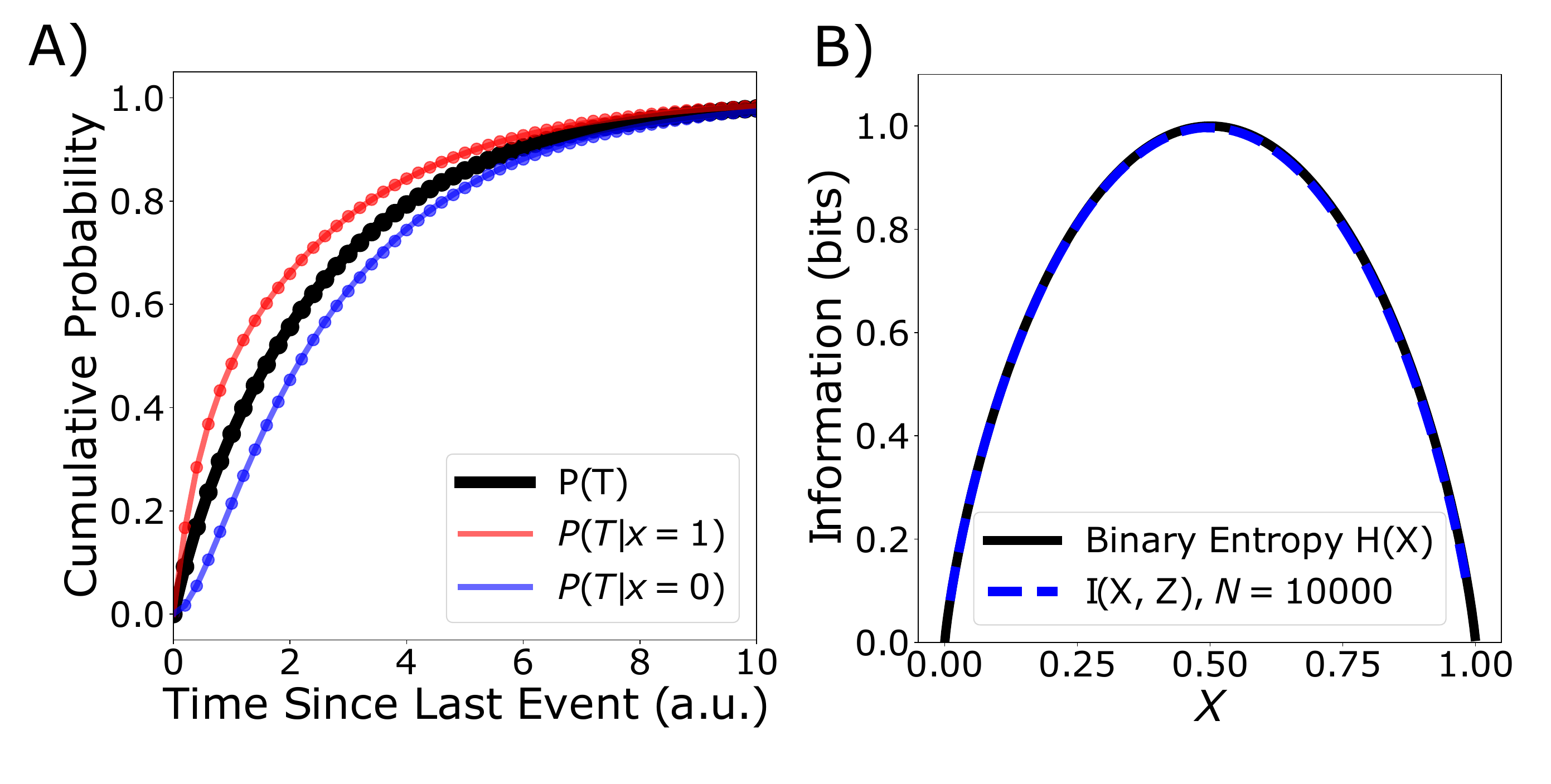}
\vspace{-1.4em}
\caption[\textbf{Comparing analytical results from solving the model defined by Eq.~\ref{EQ: definition of stochastic signal - inequality} with numerical predictions for $P(T|x)$ and $H(X)$}]{\textbf{Comparing analytical results from solving the model defined by Eq.~\ref{EQ: definition of stochastic signal - inequality} with numerical predictions for $P(T|x)$ and $H(X)$.} A) To verify the probability distribution of Eq.~\ref{EQ: Probability Time Distribution}, we ran a Gillespie simulation for Eq.~\ref{EQ: definition of stochastic signal - inequality} and sampled the system at random times to obtain the distribution $P(X, T_{\text{last event}})$. Comparing this distribution (dots) with our analytical results (line) reveals a good agreement between the theory and simulation. Data corresponds to a simulation with $k_\text{off} = k_\text{on} = \alpha = 1.0$. 
B) Plotting the maximum possible mutual information $I(X; Z)$ for $N = 100,000$ reveals that the maximum mutual information closely matches the entropy $H(X)$ at this $N$, shown by the agreement between the solid black line and dashed blue line. This suggests that when $N = 100,000$, one can infer the state of the upstream with a very high accuracy.}
\label{FIG: On-Off Switch Mutual Information Extra Figure}
\end{figure}

This yields the following probability distribution
\begin{equation}
\label{EQ: Probability Time Distribution}
\begin{split}
P(1, t) &= k_\text{on} \frac{e^{r_2 t}  - e^{r_1 t}}{r_2 - r_1} + \frac{r_2 e^{r_2 t} - r_1 e^{r_1 t}}{r_2 - r_1}
\\P(0, t) &= k_\text{off} \frac{e^{r_2 t} - e^{r_1 t}}{r_2 - r_1} 
\end{split}
\end{equation}
which is plotted in Fig.~\ref{FIG: On-Off Switch Mutual Information Extra Figure}A.

The mutual information can be computed from $I(X; Y) = H(X) + H(Y) - H(X; Y)$, where marginal and joint entropies are given by
\begin{equation}
H(X) = P_{on} \log_2 \big(P_{on} \big)
\end{equation}
\begin{equation}
H(T_{\text{last event}}) = \int_{0}^{\infty} P(t) \log \big(P(t) \big) dt
\end{equation}
\begin{equation}
H(X, T_{\text{last event}}) = \int_{0}^{\infty} P(0, t) \log \bigg(\frac{P(0, t)}{P(0) P(t)} \bigg) + P(1, t) \log \bigg(\frac{P(1, t)}{P(1) P(t)} \bigg) dt
\end{equation}
resulting in Eq.~\ref{EQ: Maximum Mutual Information}. Although there is no closed form solution to this integral, one can plot it numerically as shown in Fig.~\ref{FIG: On-Off Switch Mutual Information Extra Figure}B.

\printbibliography